\documentclass{aa}  
\usepackage{graphicx}
\usepackage{figsize}
\usepackage{txfonts}
\usepackage[usenames]{color}
 \usepackage{comment}
 
\begin{document}

   \title{The evolution of lithium in FGK dwarf stars}

   \subtitle{The Li rotation connection and the Li desert}

   \authorrunning{Llorente de Andr\'es et al.}
   \author{F. Llorente de Andr\'es
          \inst{1,2}\fnmsep\thanks{fllorente@cab.inta-csic.es},
           C. Chavero
          \inst{3},
         R. de la Reza
          \inst{4},
          S. Roca-F\`abrega
          \inst{5}
          \and
         C. Cifuentes
         \inst{2}
          }

   \institute{ Ateneo de Almagro, Secci\'on de Ciencia y Tecnolog\'ia, 13270 Almagro, Spain
      \and
   Departmento de Astrof\'isica, Centro de Astrobiolog\'ia (CAB, CSIC-INTA), ESAC Campus, Camino Bajo del Castillo s/n, 28692 Villanueva de la Cañada, Madrid, Spain
      \and
   Observatorio Astron\'omico de C\'ordoba, Universidad Nacional de C\'ordoba, Laprida 854, 5000 C\'ordoba, CONICET, Argentina
   \and
   Observatório Nacional, , Rua General Jos\'e Cristino 77, 28921-400 São Cristovão, Rio de Janeiro, RJ, Brasil
   \and
      Departamento de F\'isica de la Tierra y Astrof\'isica and IPARCOS, Facultad de Ciencias F\'isicas, Plaza Ciencias, 1, Madrid, E-28040, Spain
      }


\date{Accepted 12 August 2021}

 \abstract{
 
  We investigate two topics regarding solar mass FGK-type stars, the
lithium rotation connection (LRC) and the existence of the  ``lithium desert''. We  determine the minimum critical
rotation velocity  ($v \sin i$) related with the LRC separating slow from rapid stellar rotators, as being 5\,km\,s$^{-1}$. This value also split different stellar properties.
For the first time we explore the behaviour of the LRC for
some stellar associations with ages between 45 Myr and 120 Myr. This allows us to
study the LRC age dependence at the beginning of the general spin down stage for
low mass stars, which starts at  $\sim$ 30--40 Myr. We find that each stellar group presents
a characteristic minimum lithium (Li) depletion connected to a specific large rotation
velocity and that this minimum changes with age. For instance, this minimum changes
from $\sim$ 50\,km\,s$^{-1}$ to less than 20\,km\,s$^{-1}$ in 200 Myr. Regarding the lithium desert, it
was described as a limited region in the A(Li)-$T_{\rm eff}$ map containing no stars. Using $T_{\rm eff}$
from {\em Gaia} DR2 we detect 30 stars inside and/or near the same box defined originally
as the Li desert. Due to their intrinsic $T_{\rm eff}$ errors some of these  stars may be
inside or outside the box, implying a large probability that the box
contains several stars. Considering this last fact the  ``lithium desert'' appears to be
more a statistical distribution fluctuation than a real problem.
  }
  
   {}
   {}
   {}
   {}
   {}


   \maketitle
%

\section{Introduction}

   When the presence of the fragile atom of lithium (Li) was correctly detected and measured for the first time in the Sun by \citet{Muller1975}, a very important step was made towards the understanding of the stellar evolution of this element, $^7$Li isotope. In fact, all this evolution reflects the history of the destruction of the original lithium, with which stars were born from the interstellar matter. 
The Li abundance, A(Li), of the interstellar medium depends on metallicity  \citep{Lambert2004,Lyubimkov2016,Guiglion2019}. In view
of this, we adopted the value of A(Li) = 3.2 $\pm$ 0.1\footnote{A(Li) = log[N(Li)/N(H)] + 12}, which is also compatible with the meteoritic value of A(Li) $=$ 3.26
$\pm$ 0.05 \citep{Asplund2009}.

Stars, especially those with FGK-types and with masses
near the Solar mass, which are our concern in this work,
show two  different stages of the Li depletion in their
atmospheres. The first one happens during the rapid pre-main sequence (hereafter PMS) phase, in
which a circumstellar disk is magnetically connected to the stellar surface,
producing a halt of the stellar rotation. Because of this, a rapid initial Li
depletion occurs, in an internal region between the base of the external
convective zone (hereafter CZ) and the radiative stellar core. 
Here, an important mixing is installed \citep[see for instance the model of][]{Eggenberger2012}.

\citet{Chavero2019} applied this model to solar mass FGK-type stars, finding that with an initial original A(Li) of $\sim$
3.2, depleted values were obtained of the order of A(Li) $\sim$ 2.0 in a time scale
of $\sim$ 9 Myr, which is the lifetime of the protoplanetary disks, before stars enter into the main-sequence (hereafter MS) stage.
 During this last evolutionary phase, with a long lifetime scale of the order of
9--10 Gyr, a very slow and not well known physically Li depletion mechanism
enters into action \citep[see for instance][]{Dumont2020}. This slow depletion action was first proposed by \citet{Herbig1966}.
Nevertheless, rapid stellar rotations can interfere on the Li depletion
process. This is specially the case of the younger stellar groups, where rapid
rotators are present. An important difference then appears, in which slow
rotators are Li-poor and rapid rotators are Li-rich.

This property, which is part of the present study, is known in the literature as the ``lithium-rotation
connection'' (hereafter LRC). Since the seminal study of this effect in the Sun
by \citet{Conti1968}, the LRC mechanism has been the subject of several works. 
Contrary to some ideas that considered that a rapid stellar rotator will induce
a strong internal mixing destroying Li, works by \citet{Butler1987} in the
Pleiades open cluster (hereafter OC) with an age of $\sim$ 125 Myr and in Alpha
Per (50--70 Myr) by \citet{Balachandran88}, showed that largest rotations
preserve the Li. \citet{Soderblom1993} performed an extended research of the
LRC in the Pleiades, which was after revisited by \citet{King2009} and recently by  \citet{Barrado2016} and  \citet{Bouvier2018}.

These works confirmed and amplified the results of  \citet{Soderblom1993}. Another recent work by
 \citet{Arancibia2020} found that LRC is also present in a different type
of stellar group, as the stellar stream Pec-Eri, with  a similar age to the
Pleiades. In this context a recent important analysis in the OC M35 ($\sim$ 150
Myr) has been presented recently by \citet{Jeffries2020}. Is the LRC a
universal property in the sense that LCR is present at different ages and in
different kind of stellar groups? The LRC is present, for example, in NGC
2264, a very young OC with an age of 5 Myr \citep{Bouvier2016}, and its
presence continues up to ages similar that of the Pleiades. A recent work have found indications of the presence of the LRC in an even younger OC ($\sigma$ Ori) with an age of 3 Myr \citep{Garcia2021}.

We note that the LCR it is also observed in different stellar young moving groups or
associations \citep{daSilva2009,Messina2016}.
Several theoretical mechanisms have been proposed to explain the LRC
property. In general these mechanisms  involve the interaction between the base of the CZ
and the hot internal burning region, where Li is destroyed. These studies regard a
very slow diffusion acting in this intermediate zone  \citep{RudigerPipin2001,TschapeRudiger2001}.
Other works are related to the action of penetrative
convective plumes \citep{Siess1997,Baraffe2017}. Different
mechanisms have also been invoked consisting on the reduction of the
temperature at the base of the CZ produced by atmospheric inflation
processes  \citep{SomersPinsonneault2014,SomersStassun2017}. For a
recent complete review on the action of the LRC see \citet{Bouvier2020}.

In the present study, we first propose to determine observationally the
presence of a critical minimal velocity separating slow and rapid rotators, as
is discussed in  \citet{Baraffe2017}. Another aspect that we also  consider
here, consists on studying the LRC behaviour in stellar associations just after the
beginning of the general spin down process era for solar mass stars, which
happens at 30--40 Myr, \citep[see, for instance,][]{Bouvier2014}. 
For this, we consider three stellar moving group
candidates with a similar age of $\sim$ 45 Myr which are:
the Tucana-Horologium association (hereafter THA), the
Columba association (COA) and the Carina association
(CAA) \citep{daSilva2009}. After, we consider the AB Doradus association (ABDA) with an age of $\sim$ 120 Myr with
a similar age as the Pleiades OC with an age of 125 My
\citep{Bell2015, Bouvier2018}.

Separately from the above described research within the present work, we tackle another different research,
referring to the existence or not, of what is called the ``lithium desert'' \citep{Ramirez2012,AguileraGomez2018}.
This consists on a peculiar zone placed in the general A(Li) versus effective temperatures $T_{\rm eff}$   map, in which apparently
no stars are found.

Our work is based in a sample of  1307 field MS stars of F5 to K4 spectral types, where 244 stars contain known planets, and 1063 stars, a priori without planets. Also, we consider 265 and  151 stars belonging to
14 OC and 4 associations, respectively. We have measurements of the Li abundances for all the mentioned objects with data obtained
in the literature.

The stars chosen in this work are ``non Li-Be-B Dip'' stars
whose mechanisms of depletion, especially for Li, are related to
the shallow convection layers of these early-mid F-type stars acting in the above mentioned temperature interval 
 \citep[see for instance the model of][]{Stephens1997}. By avoiding these early-mid F-type stars we
are then considering, in general, stellar FGK objects with relatively
small masses and relative large convective layers appropriate for a
deep Li depletion acting mechanism, with stellar masses less than
1.5~$\mathcal{M}_\odot$ \citep{Pinheiro2014}.

This work is the second of a series devoted to the study of Li which started with  \citet{Chavero2019}.
This paper is organised as follows:
 Section~\ref{Sect:DataCol} presents the stellar data.
 Section \ref{Sect:3rotation} is centered on the study of the lithium-rotation
connection. In Section \ref{Sect:desert} we address the problem of the existence or not of the so called  ``lithium desert''. 
Finally, we discuss and comment  the conclusions in Section \ref{Sect:CONCLUSIONS}.

\section{Data Collection}\label{Sect:DataCol}
\subsection{Sample of field stars}

The origin of our current sample of field stars, listed in Table  \ref{t:parameters}, is our previous work, \citet{Chavero2019}.
From this first set we increase the number of objects in the sample looking for stars with Li
measurements taken with a similar methodology, and all of them come from
internally consistent ones. The search was carried out by means of using the
Spanish Virtual Observatory (SVO) tool called    {\tt VOMultiCatalog\_Interface}\footnote{http://svo2.cab.inta-csic.es}.
It looks for the demanded stellar data by scanning all the catalogues within {\tt VizieR}. In the case of lithium abundance, A(Li),  we get
information from about 11 catalogues (all references are displayed in Table~\ref{t:parameters}). Once the stars have been identified, we proceed to triage them, as explained
later.

In order to avoid any bias,  we selected our stars  without regard to their age, metallicity, rotational
velocity or any previous detection (either dust nor planets). Just to detect if there is any
influence by the presence of planets, in this work we take into account stars containing planets and
those for which no planets have been detected. We  proceeded this way since several studies
presented evidence that the planetary formation process can modify the lithium of the host star
\citep{Israelian2004,Takeda2005,Gonzalez2010}. The nature of
this disk-planet interaction will be object of study in another article.

A restrictive criterion that was imposed for the selection of the stars in the sample was, in addition
to known A(Li), that the spectral types were FGK-type stars and whose masses were M$_\star$ $<$ 1.5 $\mathcal{M}_\odot$. We
thus eliminated, from the first sample, giant and subgiant stars. We found more than 2000 stars
with reliable A(Li) values, but we dismissed this number 
of stars because we introduced a cut off in age at 10.5 Gyr. This cut is in order to eliminate the most evolved stars.

The final sample of Li abundances includes  stars from the next 11 catalogues: \citet{Chavero2019} as well as \citet{AguileraGomez2018,Bensby2018,Cutispoto2002,DelgadoMena2014,daSilva2009,Ghezzi2010,Gonzalez2015,Luck2017,Ramirez2012,Thevenin1998}.
These works were selected because their data are coming from internally consistent measurements.   
We solved the dilemma of finding values from different authors by prioritizing those from the most
recent publication. In the case of being our measured values from \citet{Chavero2019} these were
chosen in any case. For the corresponding stellar physical parameters, described below, we sought the greatest homogeneity among the different sources from which the data were collected.

The values of $T_{\rm eff}$, without any restriction or cut both upper and lower, and parallaxes, were
obtained from the {\em Gaia}  DR2 catalogue \citep{DR2Gaia2016,DR2Gaia2018}. It gives uncertainty values for the effective temperature in terms of lower (b\_$T_{\rm eff}$) and upper
limits  (B\_$T_{\rm eff}$), terms in which $T_{\rm eff}$ values appear in the  {\em Gaia} DR2 catalogue.   All our stars belong to
the ``clean'' subsample of $T_{\rm eff}$ recommended by \citet{Andrae2018}.  We obtained the values of
[Fe/H] from  \citet{Gaspar2016+} and \citet{Soubiran2016+}. $V$ magnitude values
were collected from the SIMBAD database. The search of $v \sin i$ implied searching within 30 catalogues.
In this case we established as a priority the values published by \citet{Brewer2016}, and \citet{DelgadoMena2015}.
In the case of not finding this value in the most recent works, we took them from
the catalogue of \citet{glebocki2005}. All the catalogues from which the data have been
obtained are listed in Table \ref{t:parameters}, along their respective references.

The age values have been obtained from 29 catalogues (see Table \ref{t:parameters}), in such a case we prioritised the more recent values, usually
\citet{DelgadoMena2019}. However, this age scale might introduce inhomogeneities. Thus, for
achieving greater homogeneity and for consistency with our previous work, we chose to derive ages
and other evolutionary parameters, such as masses and radii, by using the {\tt PARAM 1.3} code \citep{dasilva2006}. This code requires $T_{\rm eff}$, [Fe/H], $V$ magnitude and parallax. So we have two sources for the
age and the mass: one  that comes from  the literature, as described above, and the
other from this work using {\tt PARAM}\footnote{http://stev.oapd.inaf.it/cgi-bin/param\_1.3}.

As mentioned in the introduction, this work is part of a set of them dedicated to the study of Li,
which led us to complete the physical characteristics of our sample with the values of chromospheric
activity index, $\log R'_{HK}$, which were mainly collected from \citet{BoroSaikia2018}. When we did not
find values in this catalogue we looked for them in \citet{KrejcovaBudaj2012}. In the first case, there
are compiled 4454 stars, whose chromospheric activity was collected from a set of different sources.
Such catalogue served to study the behaviour of the variation in the chromospheric activity of cool
stars along the main sequence. The catalogue of \citet{BoroSaikia2018} listed different values for some stars.
In such a case, we catalogued the average value and, as the error, the difference
between the maximum and the minimum value. Also, using {\em Gaia} DR2 parallaxes, proper motions and
radial velocities, and using the public software package {\tt Pygaia}\footnote{http:https://github.com/agabrown/PyGaia}, we obtained the galactocentric velocities U, V, W for all stars in our main catalogue. These last kinematic parameters plus the
chromospheric activity are included in the sample  to be used in a parallel work.

An additional observation is that, even within the giant planet regime, binaries tighter than 100\,au show a different distribution of masses, suggesting a different formation mechanism and/or dynamical history \citep{Duchene2010}. In view of all these studies, we  excluded from our samples 96 binary systems with semi-major axis $a <$ 100\,au to avoid introducing any bias in our analysis. 
Table \ref{t:parameters} displays all these physical parameters and characteristics of our field sample stars with their respective reference.

  \begin{figure}
     \centering
   \includegraphics[width=7.5cm]{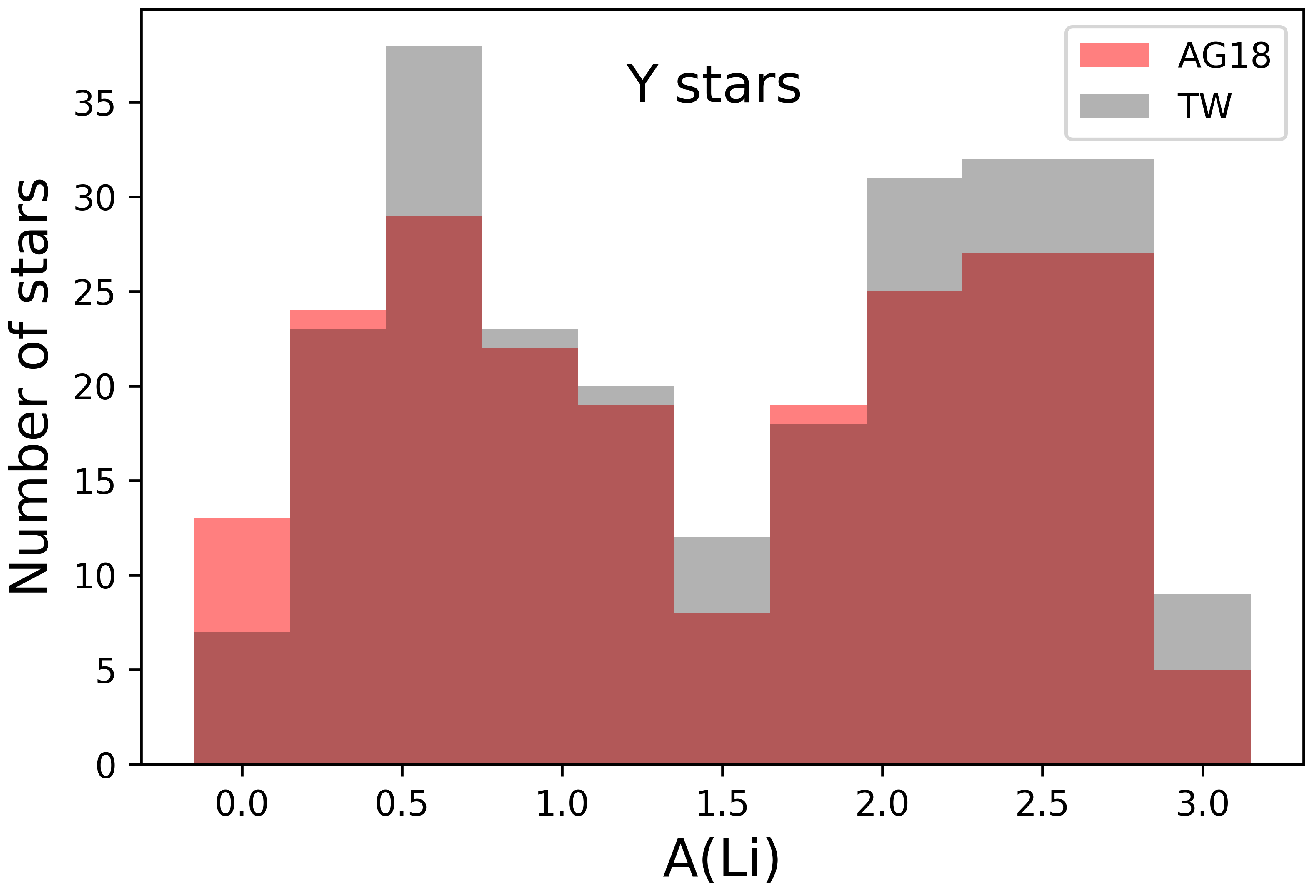}\par 
   \includegraphics[width=7.5cm]{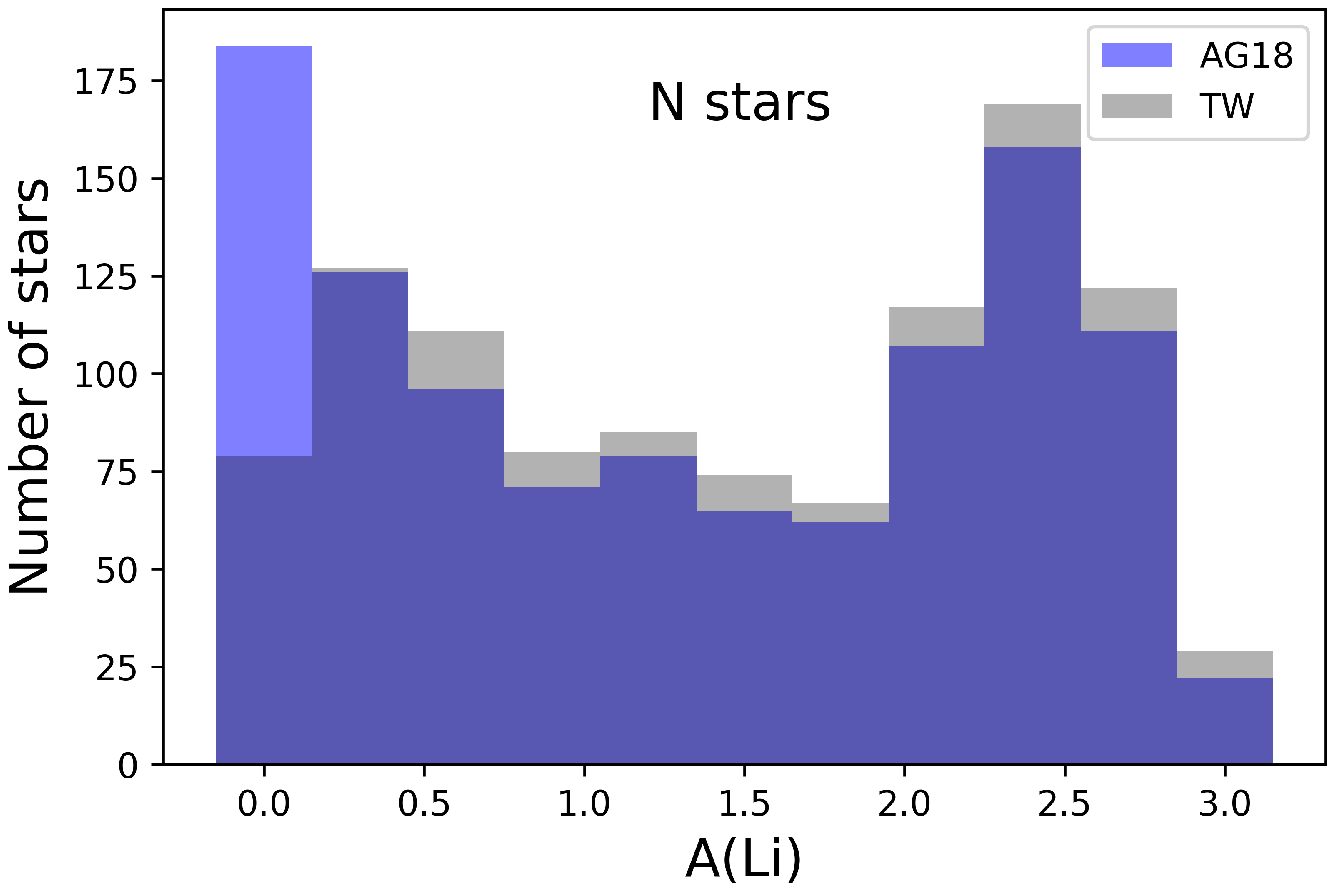}\par 
   \caption{     Statistical distribution of the sample according to the star categorisation. The histograms built from  both distributions, this work (TW) and 
   AG18, for stars with (Y) and without (N) detected planets.
}
   \label{histoAG}
    \end{figure}

\begin{center}

\begin{table*}

\caption{Stellar parameters of the  sample}
\label{t:parameters}

\scalebox{0.80}{
\begin{tabular}{lccccccccccc}
\hline
\hline
Star  & SpType   &$T_{\rm eff}$ &  M$_\star$           & [Fe/H] & A(Li) & $v \sin i$   & Age(P)  & Age(L) & $\log R'_{HK}$ &  U  V   W    & Pl\\
       &      &   (K)          & ($\mathcal{M}_\odot$) & (dex)  &       & (\,km\,s$^{-1}$) & (Gyr)   & (Gyr)  &                & (km\,s$^{-1}$) & \\
\hline      

HD1461 & G5V  & 5764 $\pm$ 81 &  1.06 $\pm$ 0.03   & 0.19 $\pm$ 0.01$^a$ & 0.6 $\pm$ 0.07$^A$  & 1.8$^1$ & 3.1 $\pm$ 2.3 & 5.6 $\pm$ 1.5$^3$ &  -5.14 $\pm$ 0.18$^4$ &  -31.74, -38.86, -1.62 & Y  \\
HD1832 &G2/3V &	5773 $\pm$ 74 &  0.98 $\pm$ 0.03   &-0.03 $\pm$ 0.03$^a$ & 1.08 $\pm$ 0.07$^B$ & 2.8 $\pm$ 0.5$^2$ & 7.3 $\pm$ 2.4 & 8.5 $\pm$ 1.4$^3$&	-4.64 $\pm$ -0.15$^4$& -7.27, -62.72, -16.94 &	N \\

...    &     &               &                  &               &             &      &              &           &        &   &   \\
\\
\hline 
\end{tabular}}
\tablebib{(1)
Col.1: Henry-Draper catalogue name; Col.2: spectral type; Col.3: effective temperature taken from {\em Gaia} DR2 where the uncertainty is taken as (B\_$T_{\rm eff}$ $-$ b\_$T_{\rm eff}$)/2; Col.4 stellar
mass calculated using {\tt PARAM} code; Col.5: stellar metallicity, (a) \citet{Gaspar2016+}; Col.6: lithium abundance, (A) \citet{Chavero2019},
(B) \citet{AguileraGomez2018}; Col.7; projected rotational velocity, (1) \citet{Brewer2016}, (2) \citet{Marsden2014} Col.8: stellar age calculated using {\it PARAM} code; Col.9: stellar age taken from literature, (3) \citet{AguileraGomez2018}; Col.10: chromospheric activity index, (4) \citet{BoroSaikia2018}; Col.11: presence of planet Yes or No; Col 12: galactocentric velocities U, V, W. Full Table  \ref{t:parameters} and references are only available in electronic format at the CDS via. 
 }
\end{table*}
\end{center}

Our catalogue is  similar to that published by  \citet{AguileraGomez2018} (hereafter AG18), but there are some differences whose consequences yield to different conclusions. So that, in order to guarantee that our data are as consistent as those already published, we  compared them with those  published by AG18, because they normalise and homogenise their final catalogue.
These authors divide in stars with (yes) and without  (no) planets. In our case, we call $Y$ stars to those with planets and $N$ stars to those without planets. 
We should know  that the AG18 catalogue is bigger than ours because it is a complete catalogue of 2318 stars, which 1470 of them are categorised in stars with or without planets. The main difference between their catalogue and ours is  the number of stars with planets (213 and 244 stars with planets, respectively), and without planets (980 and 1063 respectively). 
The mentioned difference comes from the fact that, in the sample of AG18  some stars are listed but without
mention if they host planets or not. We did not take them into account.
They are not useful for our objectives.
Moreover AG18’s sample contains a mixture of giants, sub-giants and dwarfs. In a parallel work we have already seen that the distribution of Li in giants and dwarfs is practically the same, exception made for some Li rich giants. This mixture is responsible for the difference that shows the intercept value that in the equation that relates the temperature of our sample with the sample of AG18.

 Next are the relations between  Li, [Fe/H], $T_{\rm eff}$ and age  of this present work (TW) with to respect to AG18 listed values:

\[\rm A(Li)_{TW}= 0.99 \rm A(Li)_{AG18} + 0.01, R^2 = 0.99\]
\[[\rm Fe/H]_{TW} = 0.99 \rm [Fe/H]_{AG18} + 0.001, R^2 = 0.99 \]
\[T_{\rm eff~TW} = 0.93 T_{\rm eff~AG18} +397, R^2 = 0.88\]
\[\rm Age_{TW} =  0.88 \rm Age_{AG18} +0.33, R^2 = 0.81\]

The relations are  similar, the big difference is in the number of stars with $v \sin i$ 
known: in AG18 sample, stars with planets 115 with $v \sin i$, against 245 stars in our sample; in the
case of stars without planets, the sample of AG18 contains 358, against to 931 in our sample. The two samples, AG18
and ours, show the same statistical distribution of the lithium abundance, even grouping into stars
with and without planets and the total sample size (see  Figure~\ref{histoAG}).

\subsection{Sample of open cluster stars}
\label{subSect:OCs}

To be able to have more references in terms of stars with high rotation speeds, we include stars of open clusters with an abundance of known Li that also allows us to have a larger sample of stars younger than 300 Myr.
 The aim is to establish a reference (Table~\ref{t:cluster}) constructed with the cluster data that,  in addition to the stellar sample (Table \ref{t:parameters}), help us to explain our results. The first source of data comes from the stars with Li abundances identified in the literature and belonging to open clusters, collected by \citet{SestitoRandich2005} and published in {\tt WEBDA\footnote{WEBDA database is operated at the Department of Theoretical Physics and Astrophysics of the Masaryk University.}}.  We added data of the OCs   NGC6253 \citep{Cummings2012} and NGC3680 \citep{AnthonyTwarog2009}  in order  to increase  the age range of cluster stars within the same range of  $T_{\rm eff}$ and surface gravity ($\log{g}$) that means MS stars. Ages were taken mainly from \citet{Kharchenko2013}.
 
 In the present study we assume uniform Li abundance in a
cluster before any Li depletion occurs. But, to zeroth order, the
initial Li of near-solar-metallicity open clusters should be similar
to the meteoritic abundance. 
This A(Li) value is observed in very-young open clusters, which are generally in the
range of A(Li) $=$ 3.0 to 3.4. We assumed that for the normal stars of the youngest cluster NGC
2264, plenty of T-Tauri stars, do not present Li depletion.

  Table~\ref{t:cluster} contains data of the  OCs and also young stellar associations that have been considered in this work. We have performed a deep probabilistic study to ensure that each clusters star is member of their host cluster. The membership is based upon probability following and applying the criteria quoted from \citet{Kharchenko2013} and \citet{dias2014}. So that a star is membership of its host cluster if its probability of belonging to it is around 61\% in the Dias catalogue and stars with kinematic and photometric membership probabilities higher than 60\% in the Karchenko one. With these crossed  restrictions, we admit that our selected stars are members of their respective host clusters, with a high probability of at least 90\%.
  
We  checked if the cluster stars are MS stars or not, by means of the colour-magnitude diagram (CMD) of every cluster. We conclude and confirm from all these CMD diagrams that all our sample cluster stars belong to the MS with 4000~K $<$ $T_{\rm eff}$ $<$ 6500~K. Stellar parameters of cluster stars, such as effective temperature, $T_{\rm eff}$, projected rotational velocity, $v \sin i$,  stellar mass, M$_\star$, and metallicities, [Fe/H], were collected from the literature. Exceptionally, when a $T_{\rm eff}$ value was not found we quoted the one  listed in WEBDA. The literature values are collected by means of  {\tt VOMultiCatalog-Interface}. The sample of cluster stars and their parameters is large enough  to explore how  the lithium abundance, A(Li), behaves  in stars belonging to OCs.

  The total number of remaining OC stars being part of this study is 265 objects. Their spectral types are mainly F, G and K type stars. Although the listed values of A(Li)
were computed assuming LTE (local thermodynamic equilibrium) and NLTE (non-local thermodynamic equilibrium), our study is based in general on the NLTE assumption. All the clusters
and stars data are listed in Table ~\ref{t:cluster}, which contains, for each OC/Association: identification, age
in Gyr, star name, $v \sin i$ in\,km\,s$^{-1}$,  Li abundance and  effective temperature $ T_{\rm eff} $ in K.

\begin{table}
\caption{Parameters of stars in Open Clusters and Stellar Associations}
\label{t:cluster}
\begin{tabular}{llccc}

\hline
\hline
Cluster         &  Star               &  $v \sin i$      &   A(Li)  &   $T_{\rm eff}$      \\
                &                     & (\,km\,s$^{-1}$) &          &    (K)     \\
\hline

IC 2602             &    IC 2602 120     &   51.2            &    3.19  &  5650       \\
    Age$=$0.037\,Gyr &    IC 2602 131     &   45.7            &    3.06  &  6147       \\
    (c)             &    IC 2602 132     &   14.7            &    2.98  &  6041       \\
                    &    IC 2602 157     &   11.8            &    2.92  &  5624       \\
                    &    IC 2602 2290    &   19.2            &    1.76  &  4553       \\
                    &    IC 2602 2830    &   30              &    3.22  &  5746       \\
                    &    IC 2602 2951    &   10.9            &    3.00  &  5190       \\
    ...             &   ...              &  ...              &  ....    &          \\
\hline                 
\end{tabular}
\tablebib{
(c)  \citet{Heiter2014}. Full version of the Table and references are only available in electronic format at the CDS via. }

\end{table}

 \subsection{Sample of young stellar associations}
\label{subSect:association}
 
We consider four stellar associations  and studied
for the first time  the LRC process (see Table~\ref{t:cluster})
in stars with stellar effective temperature $T_{\rm eff}$ $>$ 4000~K.
These are  three moving
groups with a similar age of $\sim$ 45 Myr, Tucana-Horologium,
Columba and Carina associations and a fourth one, the AB
Doradus association with an age of $\sim$ 120 Myr \citep{daSilva2009}. Thus the sample (field stars and OC stars) is widened by the addition of 167  stars of young stellar associations.
 
We also compare this last group to the OC Pleiades (125 Myr) because
these two stellar groups are considered to have a common 
origin   \citep{Luhman2005,Ortega2007} and they also have a similar age. These studies permit
us to see the evolution of the LRC first in an age interval of
near 100 Myr.

This study requires  a careful examination of which of the
published members of these associations    are to be consider as intruder
members. In principle, there are two
methods to detect them and these are made by means of Galactic dynamics
or by a chemical abundance analysis. In the dynamical method, the 3D orbits
of all the members of a moving group follow dynamical trajectories from an
initial point considered to be the origin of the association in the past, up to the
present observed positions. These orbits are established by the action of a
general Galactic potential. Any interloper is easily detected because its orbit
follow a very different trajectory from all the members of the group. In the
dynamical method used by \citet{Ortega2007}, used to determine the origin
of ABDA from the Pleiades, any interloper member have been detected.
Nevertheless, a chemical analysis by  \citet{Barenfeld2013} of ABDA,
suggested that some members of what they call the stream, in contrast to the
``nucleus'' of ABDA, contain certain suspicious chemical members. In any
case, the metallicity determined by \citet{Ortega2007} is compatible with that
obtained by \citet{Barenfeld2013}.

Translated into a intruder selection, we identified  as a non-member, any star that have a specific individual
metallicity value different by larger than a factor 0.1 dex from the mean
metallicity representing that of the group. For this, we  first estimate the
mean metallicity ([Fe/H]) of each association by considering the ensemble of
members presenting a homogenous distribution of metallicities.
We are aware that these are new values in the sense    
that we have not found comparative values in the literature, with
the exception of ABDA. The respective  mean metallicities are:  for THA the  [Fe/H] $=-0.04$, for COA [Fe/H] $=0.002$,
for CAA [Fe/H] $=0.02$ and for ABDA [Fe/H] $=0.02$. For ABDA, we note
that a more precise value for the metallicity obtained by \citet{Barenfeld2013} is equal to 0.02 $\pm$ 0.02. The respective intruders are not considered in
this work, and  we identify them in the Figures with an black square symbol in each of the A(Li)
versus $v \sin i$ Figures presented  and also marked in Table~\ref{t:cluster}.   

\section{On the lithium-rotation connection}
\label{Sect:3rotation}

As mentioned in the introduction, the stellar rotation plays a very important
role in the Li depletion during the PMS and MS evolutionary stages. 
 When the projected rotational velocity  is such that  $v \sin i$ $>$ $v_{\rm cr}$,
where $v_{\rm cr}$ is a critical rotation velocity the stars inhibit somehow the Li depletion, appearing as Li-rich stars.
Determining its critic value is the objective of this work. 
On the contrary,  if the $v \sin i$ $<$ $v_{\rm cr}$, it means below 5\,s$^{-1}$, the stars appear as Li-poor stars. 
This is the case for the FGK-type stars with masses
between 0.8--1.4 $\mathcal{M}_\odot$ considered in this work. There are very few exceptions that will be discussed below.
 This property represents the LRC as is defined in the introduction. In the precedent
Sections ~\ref{subSect:OCs} and ~\ref{subSect:association} we discuss the contents of our data
concerning the stellar open clusters and associations respectively.

\begin{figure}
     \centering
      \includegraphics[width=9.5cm]{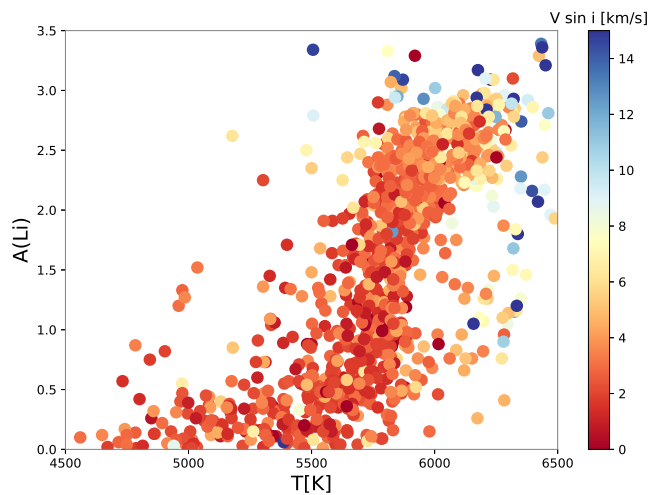}\par 
   \caption{Lithium abundance as a function of effective temperature for all stars in our catalog for our sample of field stars, where
   the color bar indicates the $v \sin i$ velocity in km\,s$^{-1}$. 
Whereas the general population is dominated by low rotating stars. 
One small group  is represented by some rapid rotators
at $T_{\rm eff}$ > 6300 K but with anomalous low A(Li) values.
}

   \label{fignewcolors}
    \end{figure}

Here, we discuss some of the main properties of field stars which
represent a large part of our data. For this we present in Figure~\ref{fignewcolors}
a general map of the distribution of the Li abundances A(Li) of
 field stars (being planet hosts or not) in function of  $T_{\rm eff}$
 {\em Gaia} values. Also,  a scale of their $v \sin i$ velocities is introduced in the
Figure. The aim of  Figure~\ref{fignewcolors} is to show the three parameters involved together, this is:
$T_{\rm eff}$, $v \sin i$ and A(Li) and to explore some indications of the presence of a global LRC
effect and also of a $v \sin i$ critical value separating rapid and slow stellar rotators. As
seen, this Figure is dominated by the presence of slow rotators (blue points) which could
be of different ages. Some medium and rapid stellar rotators (yellow and red points) at
high $T_{\rm eff}$ values are also present, which eventually suggest the presence of a vague LRC
effect. However, due to the complete mixture of ages among these field stars, and the
presence of several slow rotators stars in the Li-rich region, no robust conclusion can
 be obtained. This is also the case with the determination of any critical velocity, even
with the presence of very few stars (yellow points) with $v \sin i$~5\,km\,s$^{-1}$. We conclude
that the presence of the LRC and a critical $v \sin i$ must be studied using other
methodology and this is the subject of the next sections. 

Within the Figure~\ref{fignewcolors} are distinguished stars that do not adhere to the general behaviour. Identify themselves by presenting the following characteristics: there are 4 stars (HD149724, HD16548, HD166 and HD211080) with $T_{\rm eff}$ $<$ 5700~K and low rotation  
($v \sin i$ <~5\,km\,s$^{-1}$) but high abundance (A(Li) $>$2.0. Their high abundance is explained by their high metallicity
(average of [Fe/H] $=$ 0.22) and high activity (average of log $\log R'_{HK}$ $\sim$ -4.54 face to the mean value of -4.75) as described by
\citet{BoroSaikia2018}.

Additionally, in Figure~\ref{fignewcolors} there are six stars of late F-type with $T_{\rm eff}$>5700~K, and high rotation ($v \sin
i$>8\,km\,s$^{-1}$) but low abundance (A(Li) $<$ 1.5), these are: HD107213, HD185720, HD201203, HD30736, HD53665
HD86264. The action of rotationally induced slow
mixing can explain the low Li value as the result of slow Li mixing in these stars that are
currently undergoing angular momentum loss and are sufficiently massive (their average
mass $\sim$ 1.4$\mathcal{M}_\odot$) that interior temperatures and densities are sufficiently large to burn Li as
a result of such mixing \citep{Baugh2013}. 
More detailed conclusions on
the critical velocity will be seen in Section~\ref{Sect:criticalvelocity}. Also, many more
properties on the LRC will be obtained with the help of stars of
clusters and associations in Section~\ref{Sect:3.2}.

\subsection{ The minimum critical rotational velocity of our complete sample stars}
\label{Sect:criticalvelocity}

\begin{figure*}[!]
\centering
\SetFigLayout{3}{2}
  \subfigure[Field stars without planets]{\includegraphics{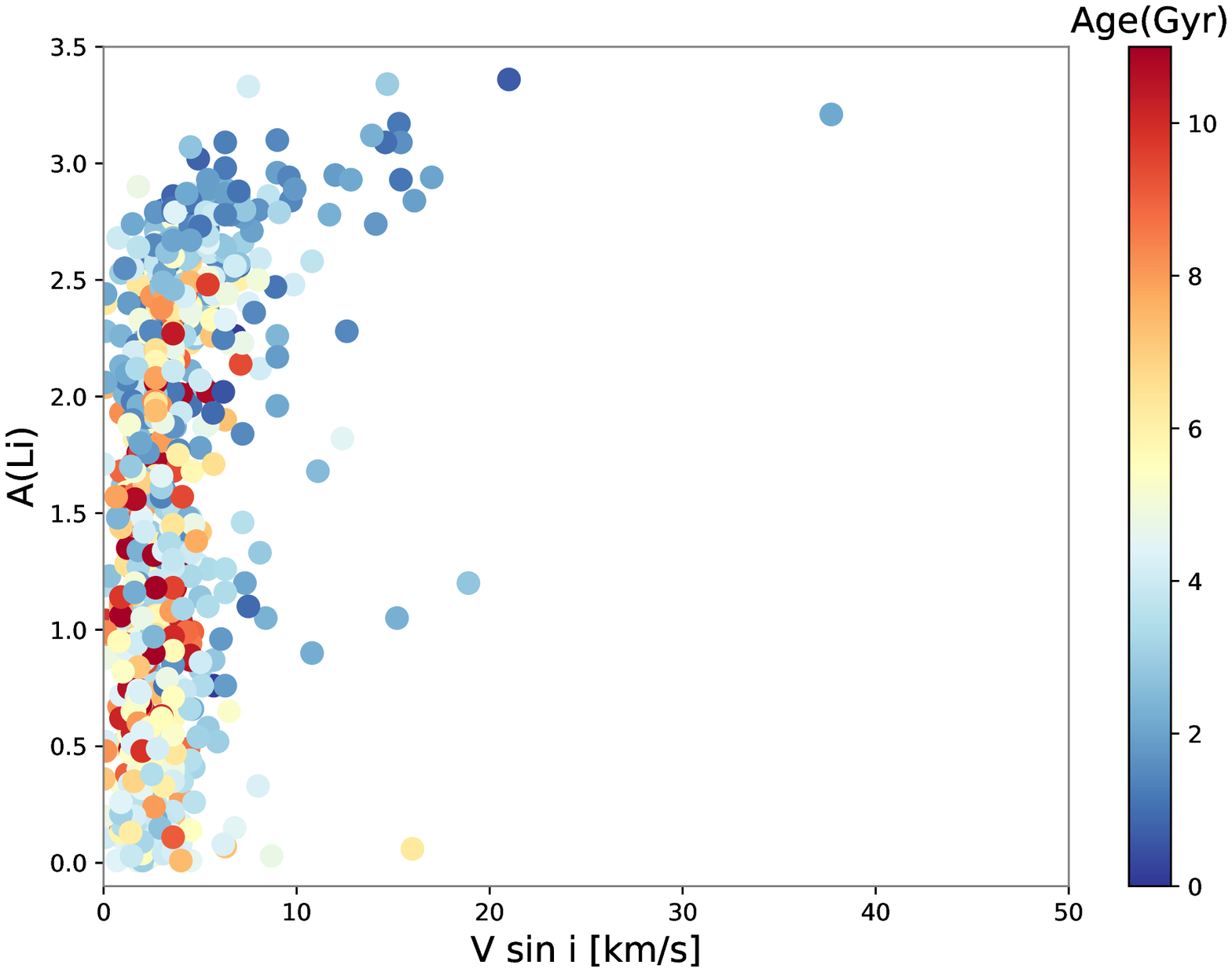}}
  \hfill
  \subfigure[Field stars without planets]{\includegraphics{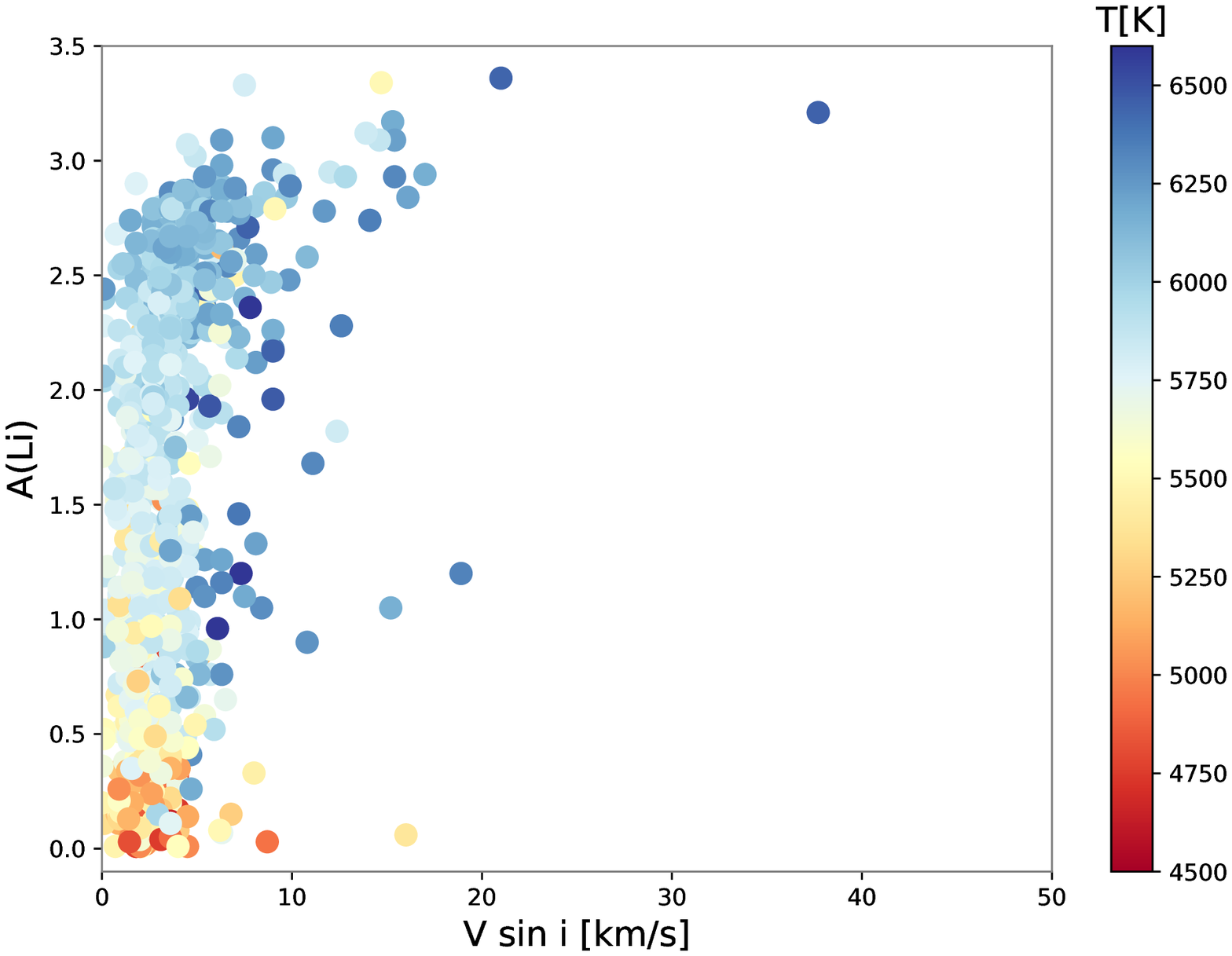}}
  \hfill
  \subfigure[Field stars with planets]{\includegraphics{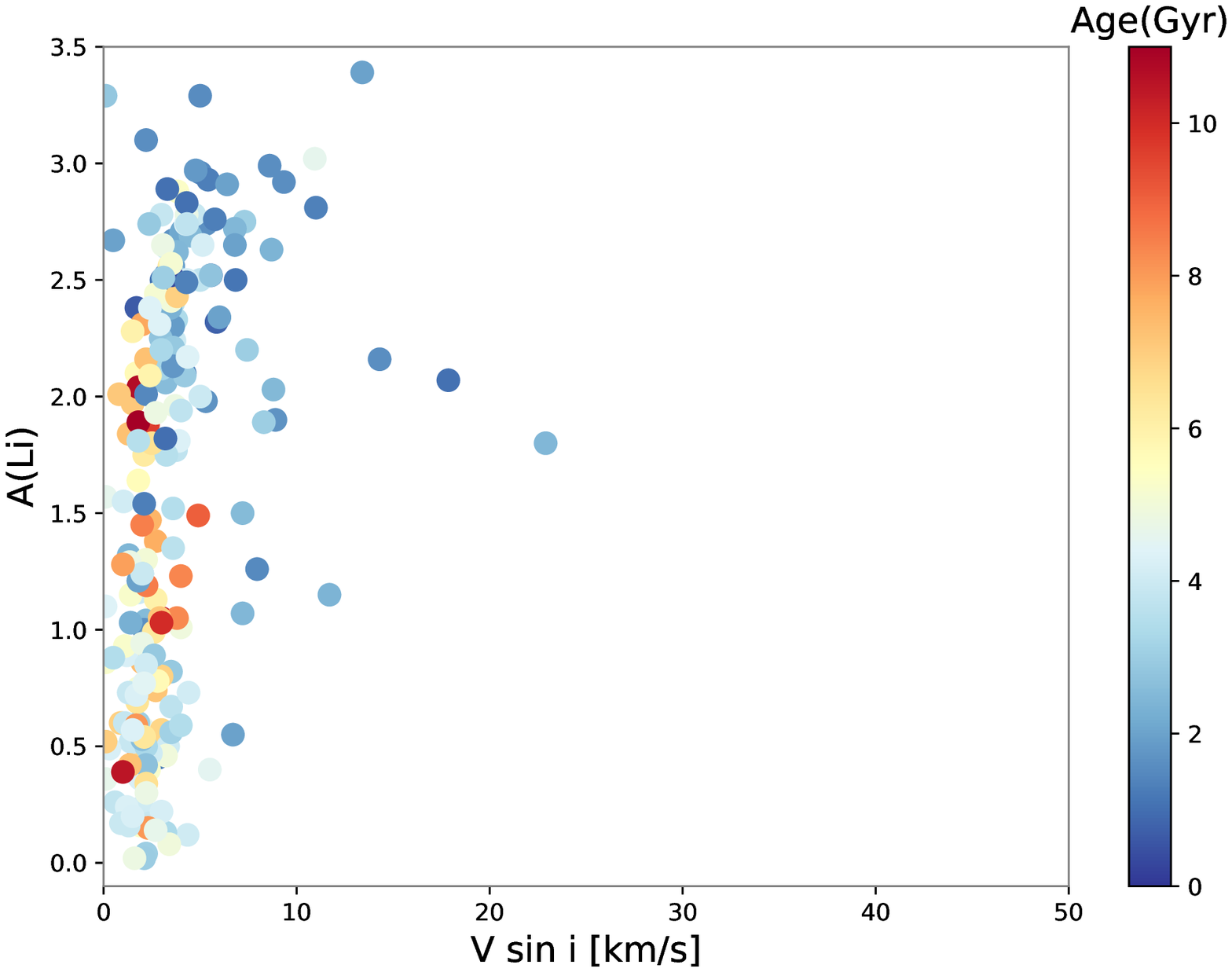}}
  \hfill
  \subfigure[Field stars with planets]{\includegraphics{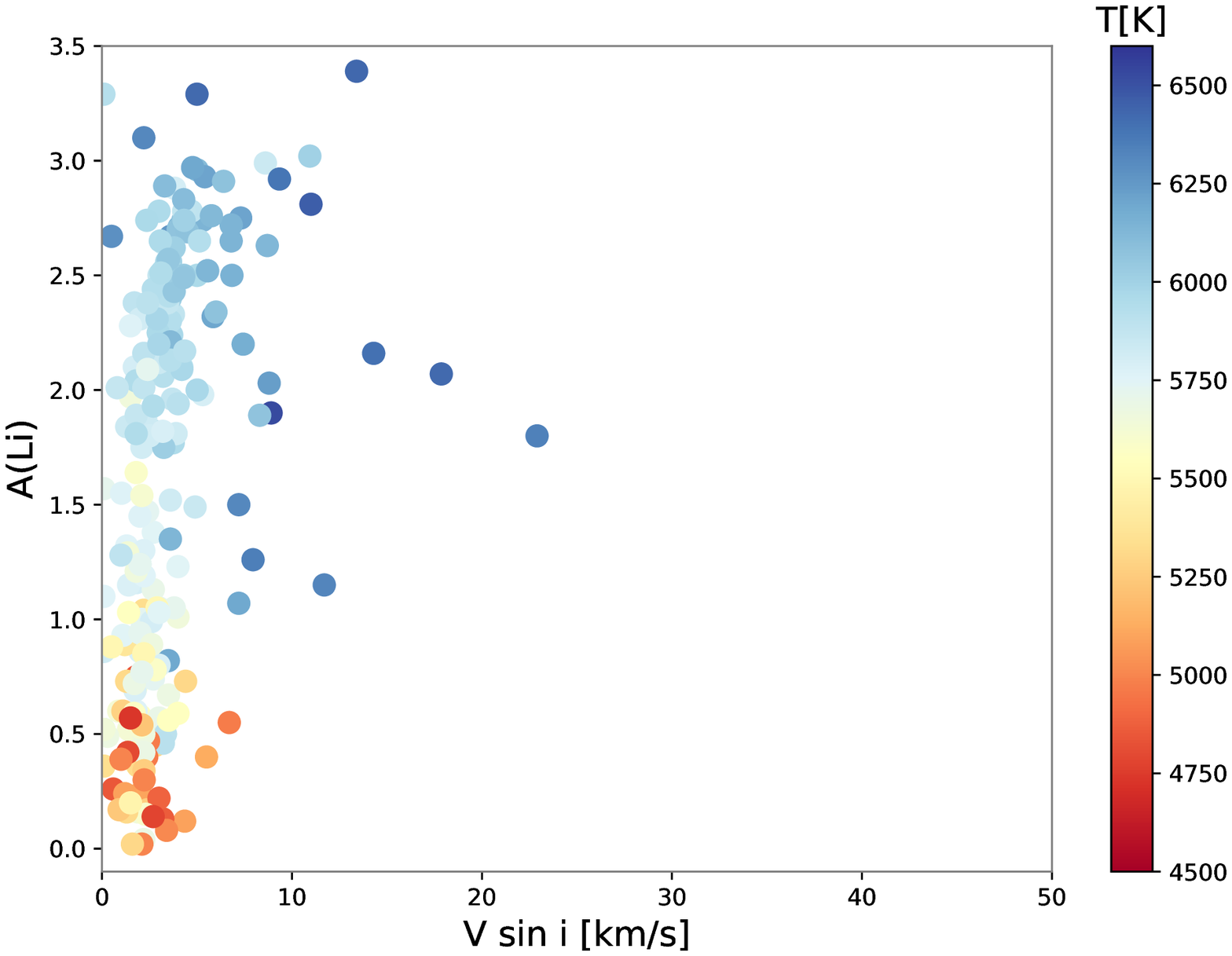}}
  \hfill
  \subfigure[Cluster stars. The age is in logarithmic scale.]{\includegraphics{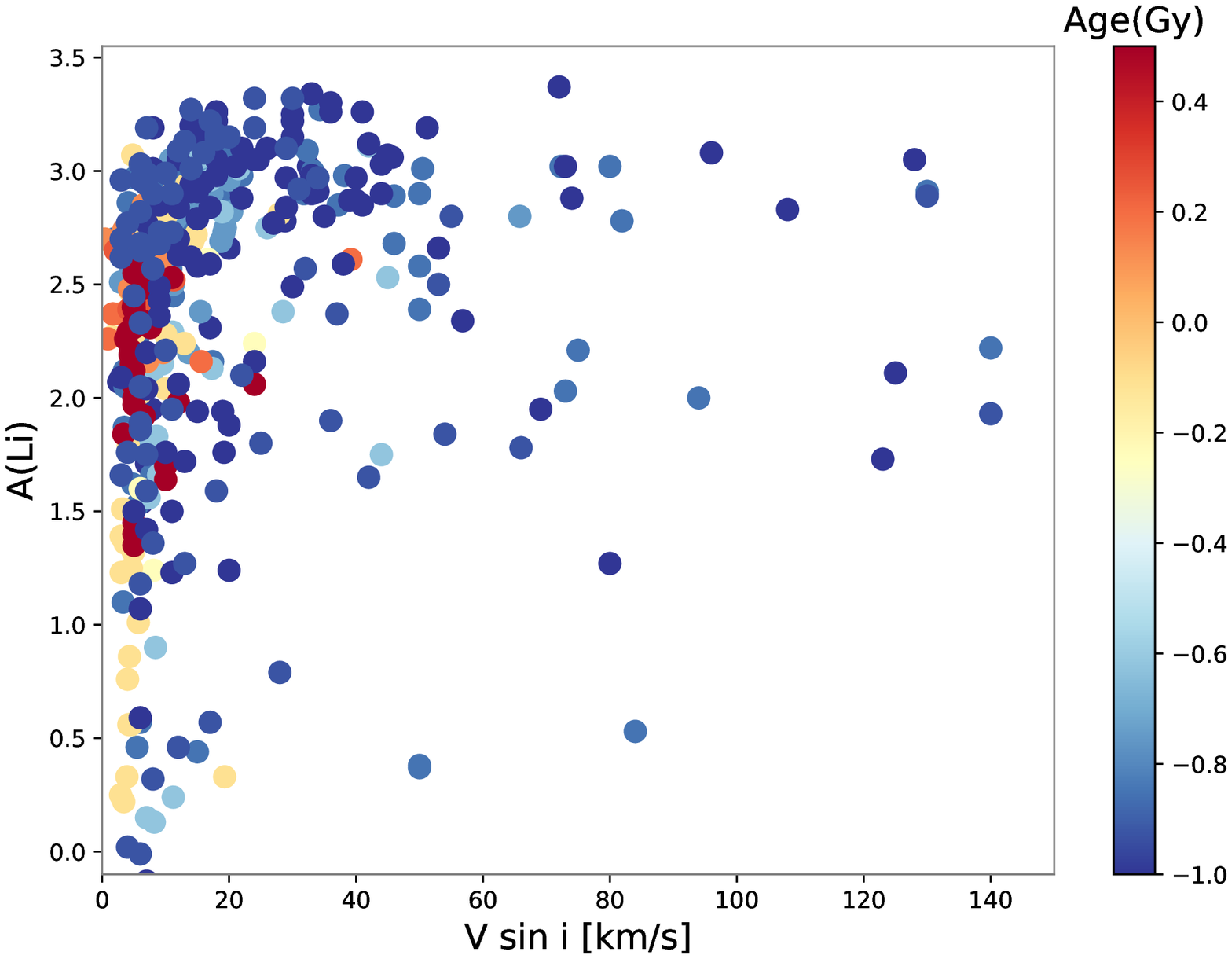}}
  \hfill  
  \subfigure[Cluster stars]{\includegraphics{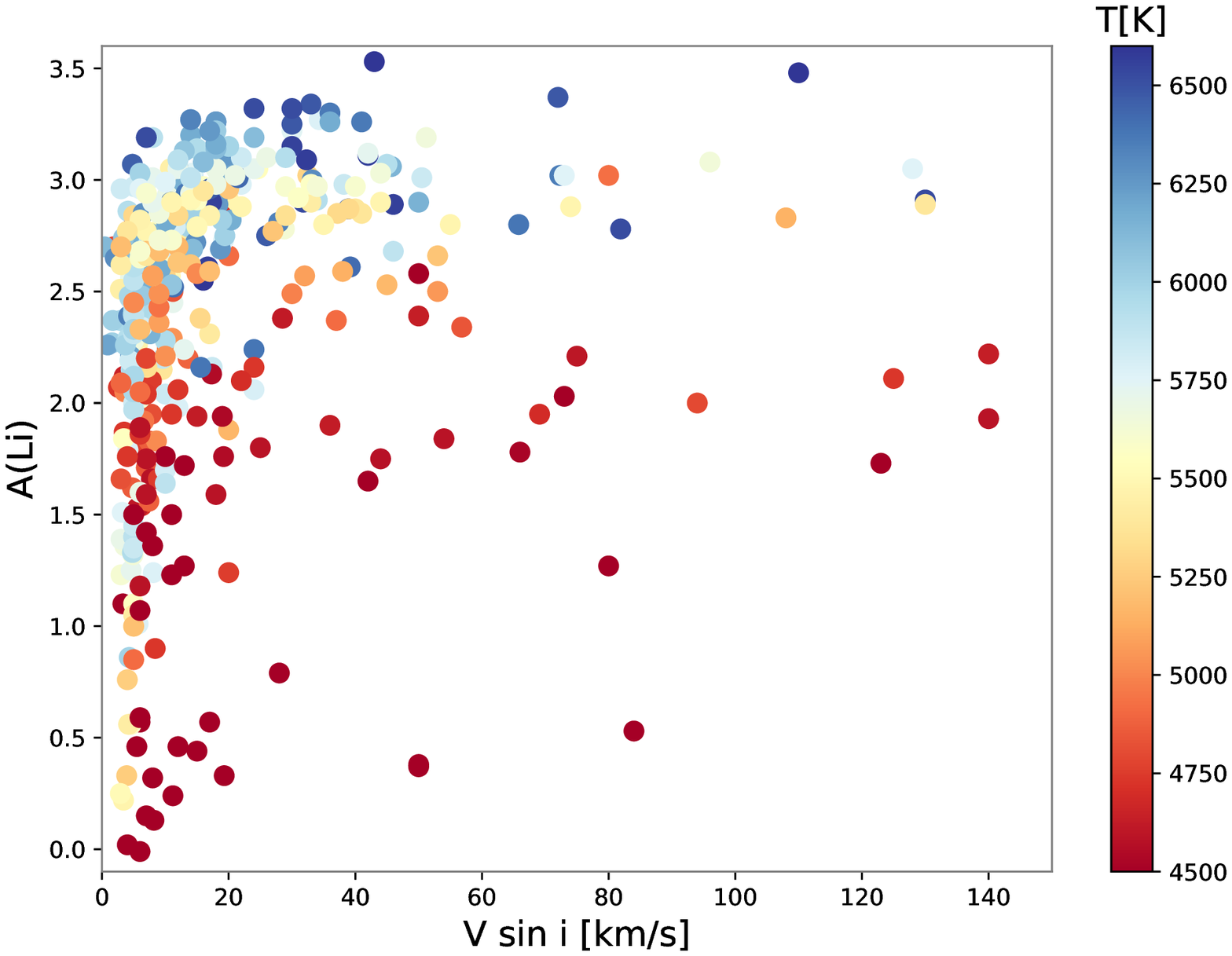}}
  \hfill
\caption{A(Li) versus $v \sin i$. Points are colored according to the  age (panels a, c, e) and the
   $T_{\rm eff}$ (panels b, d, f) of the stars as indicated by the color bar on the right of the graph.
 Panels (a) and (b) corresponding to field stars without detected planets, whereas panels (c) and (d) corresponding to field stars with detected planets.
 Cluster stars are represented in the panels (e) and (f).  The diagram  of panel e)  shows the age in 
a logarithmic colour scale ranging from dark blue for 0.035 Gyr to red for 5 Gyr. }
 \label{fig3}
\end{figure*}

 
In this section we set out to obtain a numerical figure that is the actual physical value, both for field  and cluster stars.
Figure~\ref{fignewcolors} guides us to visualise a value for critical rotational velocity, separating fast and slow rotating stars. 
Figure~\ref{fig3}   shows the A(Li)  as a function of the projected rotational velocity, $v \sin i$,
colour-coding with the age and the effective temperature.  We must note that the ages estimated for field
stars (see Table~\ref{t:parameters}) are much less precise than those of cluster ages (Table~\ref{t:cluster}). We have not represented the masses on the third axis since we only have the values of these for the field stars, in doing so with the $T_{\rm eff}$, we preserve homogeneity, while it is a way of representing the equivalent masses. In short, 
the identification  of the critical rotational velocity involve also the other involved stellar parameters.

Our search for this critical$v \sin i$ rotational velocity, involves also
the investigation of the effects of the physical parameters mentioned above.
Our method was based on analysing the mosaic depicted in Figure~\ref{fig3}. We first consider the case of the very slow rotational velocities with
$v \sin i$ $\leq$ 5~km\,s$^{-1}$.
In Figure~\ref{fig3}, for the whole lithium abundances ranging A(Li)$\sim$ 0.0 up to 3.3,
the field and cluster stars present a mixed collection of ages and $T_{\rm eff}$
values. This collection  represents some different Li evolution scenarios. For
field stars without planets (Figure~\ref{fig3} a, b) and with planets (Figure~\ref{fig3} c, d), 
we can distinguish two intervals of A(Li) bounded by the value of A(Li) $\sim$ 2.5.
In general, stars with A(Li) $<$ 2.5 are cooler than $\sim$ 6000 K and older than 4 Gyr. For the
less numerous Li-rich stars with A(Li) $>$ 2.5, these are younger than $\sim$ 3 Gyr and
hotter than 6000~K. We emphasise that this behaviour is similar for stars hosting
or not planets. In the case of clusters (Figure~\ref{fig3}e, f) the presence of quite very cool
stars ($\sim$ 4500 K) is more accentuated for A(Li) $<$ 2.5. For Li-rich stars with A(Li) $>$ 2.5, stars appear to have temperatures of 6000~K or higher, with ages between $\sim$ 0.2 Gyr and 0.6 Gyr.
We must note that the separating value of the lithium
abundance of 2.5, not only distinguish two different zones of age and temperatures as can
see in Figure~\ref{fig3}, but, remarkably, is similar to the upper limit of the final PMS depleted Li
values between 2.2 and 2.4 as found in \citet{Chavero2019}.

We also examine the situation for larger rotational velocities ($v \sin i$ $>$ 5~km\,s$^{-1}$)
for field stars, hosting or not planets, and for those belonging to clusters. First
of all, stars  with A(Li) $>$ 2.5  present, in all cases,  a normal
behaviour for the LRC phenomena. This means, for high rotation velocities,
stars maintain in general their original lithium abundance. We note that cluster
stars, differently from  field stars, present very high rotational velocities
as expected because some of them are younger than those of our field sample.
What appears to be more peculiar in this case of larger velocities, is the presence
of dispersed, relatively Li-poor stars with A(Li) $<$ 2.0. This peculiarity is more
notorious if we consider that the temperatures of field stars dispersed are different and
opposite than those in clusters. Dispersed field stars are hot and cluster stars are
cool. However, in both cases they are relatively young objects. This difference
indicates different origins. In the following we try to explain these origins.
We first consider the case of the OC.

A complete discussion on this is given in the next Sub-section 3.2. Nevertheless, we can say
that in OC, which occupy a large range of ages, their stars suffer after 30-40
Myr a general spin down effect \citep{Bouvier2014}. During this process, the cool K-type member stars are the first, 
in each cluster, to be Li depleted. In this way, 
from ages around 100 Myr and older ages, K-type dwarfs began to
successively down to the low Li abundance zone (see for example the case of the
Pleiades OC with an age of 125 Myr in Fig. 5 (b). The presence of the dispersed
cool stars and relative young stars in clusters in Figure~\ref{fig3}(e, f) is the result of this
process.

The distribution of field stars as shown in Figure~\ref{fig3}(a to d), requires a different
explanation from that of the OC, this because their evolution differ from what
we call, the general and standard Li depletion processes in solar type stars. As
mentioned before, the most important Li depletion occurs in the
initial T Tauri phase. Here, around 70 \% of T Tauri stars \citep{Armitage2003}
maintain their accretion disks with lifetimes up to 10 Myr. In the introduction
we mentioned that using the accretion disk braking model of \cite{Eggenberger2012}, the Li abundances are reduced to A(Li) values near 2.0 
\citep{Chavero2019}.

In Figure~\ref{fig3} (a to d) two different zones appear to escape this standard Li
depleting process. They are represented first, by few Li-rich objects, with low
rotation velocities ($v \sin i$ $\leq$ 5~km\,s$^{-1}$). The second group is formed by dispersed Li-
poor objects with more larger velocities ($v \sin i$ $>$ 5~km\,s$^{-1}$). These stars are hotter than 6000~K and younger than 4~Gyr. Summarising, these last mentioned
two groups present mechanisms inverse to those acting in the standard Li
process. Even if these two mentioned groups represent minor groups, they
require eventual future, non standard physical explanations, which consider
supplementary mechanisms in the complex disk-star interaction \citep{Ireland2021}.
We  presented an ensemble of different stellar properties, as age and
temperatures (or mass equivalent) for stars with rotational velocities less or
larger than $v \sin i$ $=$ 5~km\,s$^{-1}$. Considering all these physical evidences, we can
conclude that the velocity of $v \sin i$ of 5~km\,s$^{-1}$ is the most representative critical
rotational velocity to separate, not only slow and rapid stellar rotators, but  also differentiated stellar properties

This  determined critical value of $v \sin i$ is important in the
light of models trying to explore the LRC mechanism. For instance
\citet{Baraffe2017} used for this a new diffusion mechanism by means
of hydrodynamic simulations. They introduce an overshoot in the form
of plumes at the lower edge of the CZ. Their results show that the stellar
rotation affects the mixing in the CZ bottom. One of their main results, is
the prediction of the existence of a critical rotation above which the
rotation prevents the penetration of any plumes. In addition, below this
critical velocity value, rotation has small or no effects, even in the case
of most vigorous plumes.

\subsection{ The complete pattern of the lithium rotation connection    }
\label{Sect:3.2}

The representation of the variation of the A(Li) in function of
rotational  $v \sin i$ velocities for a stellar group (OC or association) is a very
practical representation to distinguish changes in the
behaviour of the Li depletion. The general pattern of the LRC is formed by
two very different behaviours.
On the one hand, by the presence of stars with
very small $v \sin i$ values ($\lesssim$~5\,km\,s$^{-1}$) covering   a wide range of the A(Li)
depleted values, from the initial value of $\sim$ 3.3 down to very depleted near zero
values in some cases.
It is generally considered that stars presenting this
behaviour have braked their larger past rotations,  for example by the PMS
mechanism as mentioned in the introduction \citep{Eggenberger2012} and
by a slowly Li depletion mechanism during the MS. On the other hand, a
complete different behaviour appears when the projected rotational is over the initial
 threshold velocity of $\sim$~5-10\,km\,s$^{-1}$.

These stars appear to be less
and less Li depleted. This pattern shows that the Li depletion is reduced
progressively up to a minimum stage. As we will see later, this specific
minimum Li depletion changes with age. Let us reconsider and ask ourselves if the Li depletion for all
large and very large rotational velocities continues to be constant or this one
diminishes. Is this behaviour age dependent?

Thus Figure~\ref{fig3}b shows that there are some speeds above 5\,km\,s$^{-1}$ at which the A(Li)
reaches the primordial value to decrease for larger $v \sin i$ values.
Not being able to know precisely what these critical speeds were, we decided to examine
here the behaviour of the LRC action using young stellar
moving groups or associations with different ages.
These are:  the Tucana-Horologium, the Columba   and the Carina associations with a similar age of 45 Myr \citep{daSilva2009}, and  the AB Doradus association with an age of $\sim$~120 Myr \citep{Bell2015}. We also examine the Pleiades OC.

The first indication of the presence of the LRC action for a representative number of
moving groups or associations, can be found in \citet{daSilva2009}. This was made for an ensemble of nine associations which
present different degrees of the LRC effect. However, no association has
been studied in detail with the exception of the young (20 Myr) Beta Pic
association \citep{Messina2016}. This last work was made by means of
observed photometric periods detecting however, that the LRC effect 
appears to be more important, only for lower mass stars with masses in the interval of 0.3-0.8$\mathcal{M}_\odot$.

\subsubsection{The Tucana-Horologium, Columba and Carina associations}

\begin{figure}
     \centering
   \includegraphics[width=8.5cm]{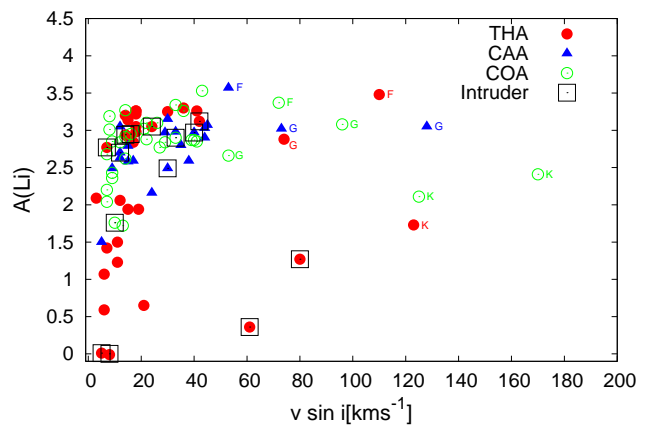}\par 
   \caption{A(Li) versus $v \sin i$ of genuine stars of the associations Tuc-Hor,
Columba and Carina. Different symbols represent: red filled circles for Tuc-Hor, green open circles
for Columba and blue filled triangles for Carina. Intrude stars not considered in this work,
are shown with a black square overlapped to the different symbols. All A(Li) values are maintained as presented by the authors \citep{daSilva2009}. 
The spectral types are marked for rapid stellar rotators.}

   \label{fig3Asso}
    \end{figure}

The stellar moving groups or associations THA, COA and CAA are sparse
groups of stars located at distances between 35 and 160~pc. Each
association is distinguishable by their similar space velocities in all of its
components. Each association is considered to have a specific spatial
formation region in an interstellar cloud, now vanished. It is from this original
place of formation that all members began to move. The time elapsed since
the considered epoch of formation and the time of the present observable
positions, is one of the methods to obtain the age of the group. The discovery
of THA has been made by \citet{Torres2000} and \citet{ZuckermanWebb2000},
whereas the discovery of COA and CAA were made and discussed by \citet{Torres2008}.
The source of A(Li) and $v \sin i$ data to study the LRC for these
three moving groups are taken from \citet{daSilva2009}. In Figure ~\ref{fig3Asso} we present the
distribution of A(Li) in function of $v \sin i$  for the ensemble of these
three associations having similar ages, by different colours. The internal
errors of the A(Li) values have been discussed in \citet{daSilva2009} and are in general
less than a factor 0.2. In any case the internal errors will not modify in a
significant way, the general distribution of points in Figure ~\ref{fig3Asso}. The presented
values of A(Li) and $v \sin i$ are then sufficient to reach our goal. The rotational
velocities $v \sin i$ are projected velocities that do not represent always the true
equatorial rotational velocities. Some considerations about these differences
of velocities will be discussed later in Section~\ref{Sect:3.2.3}.

 The respective number of our detected intruders members in the 
lists of  \citet{daSilva2009} of these three associations are: THA (8 intruders), COA (3) and CAA (3). All these considered non-members are shown in Figure~\ref{fig3Asso} marked with a black square overlapped to the different symbols 
corresponding to each association. We adopt here the ages of \citet{Bell2015}. Using a self-consistent isochronal scale they obtained the following ages for these three associations: THA (45  $\pm$ 4 Myr);  COA($42^{+6}_{-4}$~Myr) and CAA($45^{+11}_{-7}$~Myr). For our comparative analysis we
consider a mean age equal to 45 Myr  for all of them.

The distribution of points
in Figure~\ref{fig3Asso} representing the real considered members of THA, COA and CAA
have the general typical pattern expected to reflect the LRC as mentioned
before. The LRC is represented by stars showing highly Li depleted stars for $v \sin i$ $<$ 10\,km\,s$^{-1}$.

For larger $v \sin i$ velocities, an important feature appears in Figure~\ref{fig3Asso}. This is the presence of a minimum of the Li depletion corresponding to the rotational velocity of $v \sin i$ $\sim$  50\,km\,s$^{-1}$. For even larger velocities, FG-type stars are distributed in an almost horizontal sequence. Contrary, K-type dwarf stars shown a much more important Li depletion for very rapid stellar rotators. 
 As we explain in the next subsection, older
stellar groups present different behaviours regarding this general stellar
distribution of the LRC.

\subsubsection{The AB Doradus association}

\begin{figure}
     \centering
   \includegraphics[width=8.5cm]{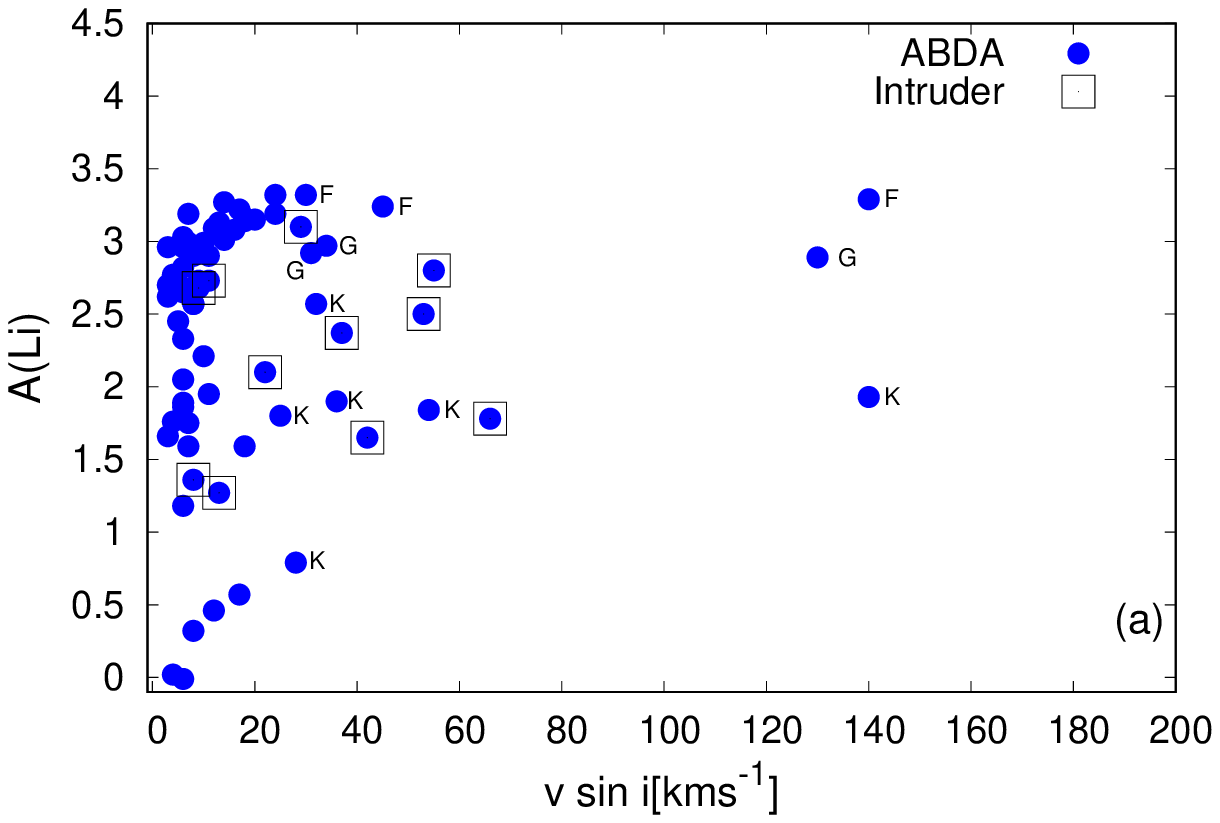}\par 
     \includegraphics[width=8.5cm]{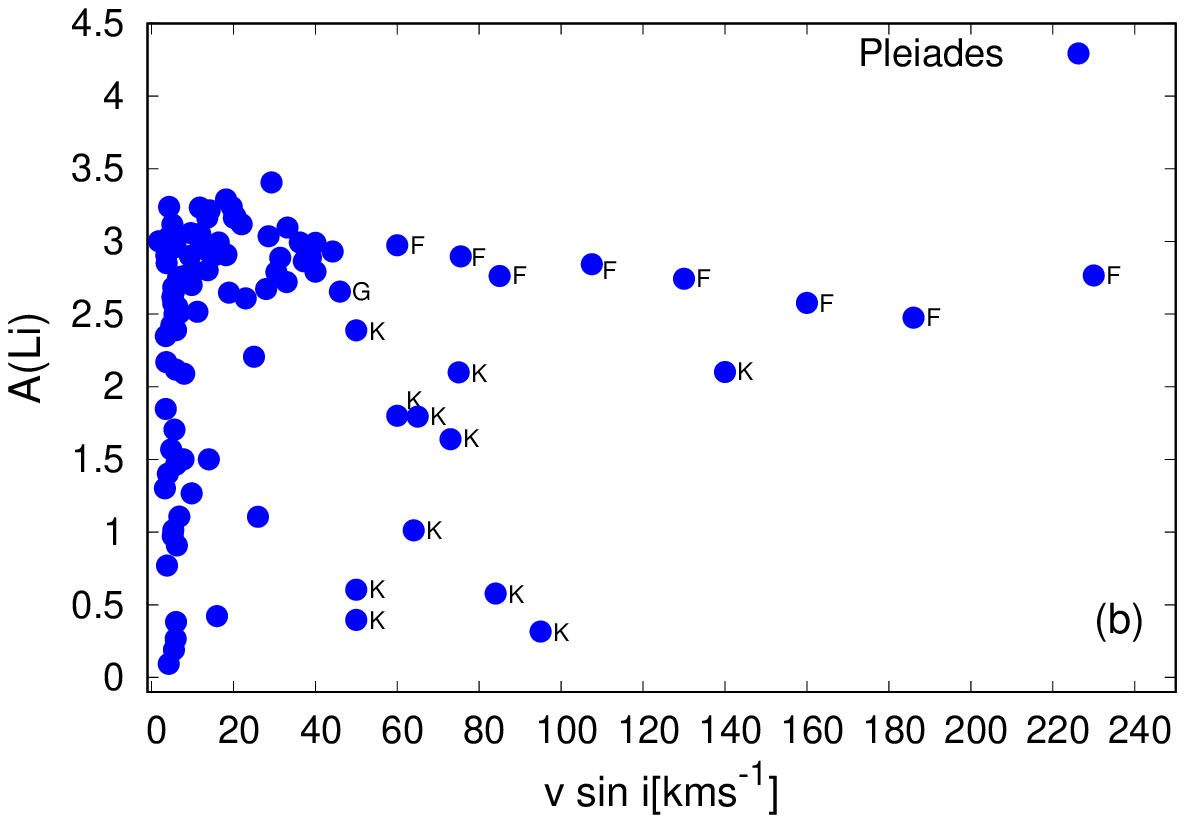}\par 
   \caption{Same representation as in Figure~\ref{fig3Asso} for stars of the AB Dor
association with an age of $\sim$ 120 Myr (top panel a) and of the open cluster
Pleiades with an age of 125 Myr (bottom panel b).
 As in Figure~\ref{fig3Asso}, the distribution shows that the few more Li depleted
stars are K-dwarf stars. The spectral types are marked for stars with $v \sin i$ $>$ 30. Intruder stars in AB Dor are represented by open
squares. Values of the Pleiades stars are taken from \citet{Barrado2016} }

   \label{fig4pleiades}
    \end{figure}

The richest populated association ABDA is the most studied among the
moving groups considered here. The literature gives a collection of different
ages. They are in an increasing age scale from 50--70 Myr \citep{Zuckerman2004,Torres2008,daSilva2009} to a coeval age
of 119 $\pm$ 20 Myr in \citet{Ortega2007}. More recent age values for ABDA
and lastly, that of \citet{Bell2015}, which gives an age of $149^{+51}_{-19}$~Myr.
In order to study more in detail the common origin of ABDA and
the Pleiades OC we  compare the action of the LRC in both stellar
groups.  Figure \ref{fig4pleiades}a shows the distribution of A(Li) in function of $v \sin i$
velocities of stars of ABDA. Figure \ref{fig4pleiades}b presents the same distribution for stars
in the Pleiades for which both values of the  $v \sin i$ projected rotation velocities and the
A(Li) values are took from \citet{Barrado2016}. New features appear for
these stellar structures, almost 100 Myr older than the previous three younger
associations considered. In ABDA the minimum Li depletion appears at  $v \sin i$
$\sim$30\,km\,s$^{-1}$, which is a smaller value than that of $\sim$ 50\,km\,s$^{-1}$  of the 45 Myr
associations. At this velocity, two branches of stars appear and instead of one branch  as was the case for the younger
are those of \citet{Barenfeld2013}, which established a minimum age of 110 Myr (Figure \ref{fig3Asso}). 

As it can be seen in both Figures \ref{fig4pleiades}a and \ref{fig4pleiades}b, the general pattern is similar. A first branch showing very strong Li depletions
appears, specially for the case of the Pleiades. This branch initially formed 
appears at $v \sin i$ $\sim$  30\,km\,s$^{-1}$ is then followed by very Li depleted stars, all of them
represented by K-type dwarf stars. A second branch, almost horizontal, with
much less Li depleted values appears, formed essentially by F-type stars .
However, in this  ``horizontal'' branch, some examples of rare, very fast
rotators G and K-type stars are present, indicating a normal stratification of
some Li depleted values. In the next sub-section we discuss more in detail
this effect.

\subsubsection{On the nature of some very rapid rotators in associations:
Projected versus equatorial rotations velocities}
\label{Sect:3.2.3}    
    
    An important work that helps to understand the effects of rotation in
associations is that of \citet{Messina2010}.
For these associations they measured the photometric stellar rotation
periods and, at the same time, they estimated the stellar radii by comparing
the stars positions in a colour-magnitude diagram with evolutionary tracks.
This data enabled them to calculate the equatorial rotation velocities for the
individual stars as $\mathcal V_{eq} = 2 \pi (\rm R/\rm P)$ where $\rm R$ is the stellar radius and $\rm P$ the
rotation period. Figure 5 in \citet{Messina2010} represents a diagram of observed
projected $v \sin i$ velocities versus the calculated and measured $\mathcal V_{eq}$ velocities.
It is interesting to note that in general, for $v \sin i$ or $\mathcal V_{eq}$ values
less than $\sim$ 50\,km\,s$^{-1}$, a large part of the individual points pertaining to several
associations, are placed near the diagonal defined by $v \sin i$ $=$ $\mathcal V_{eq}$, corresponding to equator-on orientation.
This Figure shows however, the existence of a fraction of stars with $\mathcal V_{eq}$ velocities larger than the corresponding $v \sin i$ values.
In other words, the projected observed $v \sin i$ velocities sub-estimate the real
$\mathcal V_{eq}$ velocities. Because stellar radii are fixed, this means that some stars are
rotating faster due to smaller periods, than would be indicated by the $v \sin i$
values. 

\begin{figure}
     \centering
   \includegraphics[width=8.5cm]{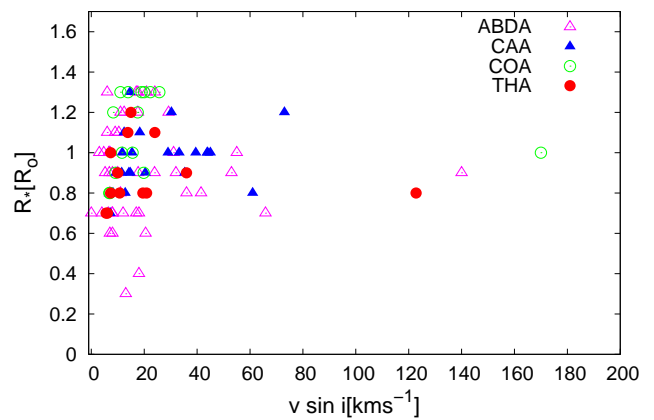}\par 
   \caption{Stellar radii versus  $v \sin i$ for
the associations Tuc-Hor, Columba, Carina and AB Dor.  It can be seen that the radii of the more
rapid rotators corresponds to the radii of K-type dwarf stars which are those
presenting more Li depleted values.}

   \label{fig5NEW}
    \end{figure}

In Figure~\ref{fig5NEW}  we present a different type of representation
that  help us to distinguish among the faster rotators, which types of stars
deplete their Li more rapidly. In this Figure the estimated stellar radii taken from  \citet{Messina2010}
for associations ABDA, THA, COA and CAA  are presented 
against the projected $v \sin i$ observed values. For larger values of $v \sin i$ (equivalent to real larger $\mathcal V_{eq}$ values)
 the corresponding stellar radii are between
0.8 and 1.0 $\mathcal{R}_\odot$, which represents well the K-type dwarf stars radii values.
These stars normally  deplete their Li faster than
FG-type stars in the rapid rotators zone due to their larger CZ.

\section{On the lithium desert}
\label{Sect:desert}

The existence of a specific and a relatively small region in the general A(Li)-$T_{\rm eff}$ stellar map,
apparently devoid of field stars, was proposed by
\citet{Ramirez2012} (thereafter RA12) and was called the ``lithium
desert'', defined by the interval 5950 $<$ $T_{\rm eff}$ $<$ 6100\,K  and 1.55 $<$ A(Li) $<$ 2.05.
The existence of such a region was inspired by the work of
\citet{Chen2001}, which found an empty area centered at A(Li) $\sim$ 1.5. This zone was separated by a bi-modal distribution of Li-rich and Li-poor stars.
They considered  that due to a strong correlation
found among low Li stars of the Li Dip stars between mass and [Fe/H] by
\citet{Chen2001}  that could also be the case in the low Li side of
the empty area. In this sense, according to this, the low Li side could contain evolved Li dip stars. 

This apparent connection with Li dip stars was discussed in RA12,
indicating that a relation between mass and [Fe/H] in the higher Li side
also exists, and this was found using the same data of \citet{Chen2001}.
At the same time RA12 found, this time using their own data, that a
similar mass-metallicity relation exists in the lower Li zone, but with a
somewhat larger dispersion. All these properties led RA12 to consider
that the contribution of evolved Li dip stars might not be  important.

Nevertheless, the main problem that appeared  consisted on
answering the question of what causes this Li desert. A process of rapid
and not yet identified Li depletion mechanism could then exist in order to
explain the complete absence of stars in this intermediate void
region. A more recent research by   \citet{AguileraGomez2018}
(hereafter AG18) using  twice the number of stars in RA12,
studied in detail the problem of the lithium desert. They proposed that
both scenarios (evolved Li Dip population and a severely Li depleted
zone) were partially correct. In any case, in both works
(RA12 and AG18), the problem on how to explain  the physical reason of this apparent
absence of stars in the Li desert remained.  We note, however, that AG18
found two apparently normal  stars (one without planets, HD 90422,
and another  with planets, HD 31253) located in the box of the Li
desert.

Not only AG18, but there are other authors
who have found the presence of stars inside that box, for example \citet{LopezValdivia2015}, who detected three stars inside the lithium desert:
BD+47, HD 44985 and \#58440 of the OC M4 ([MVB2012] 58440). 
Consequently, it seems that the Li desert is not completely
empty of stars. In fact this hypothesis became a reality, because as a result, we identify that several stars could be present in a box
(hereafter Box) corresponding to the Li desert. 
We detect some ``new'' stars
belonging to the Box  by the means of the use  of new $T_{\rm eff}$ values
furnished by space observatories as  {\em Gaia}.

In a certain way, this is a challenging problem due to the reduced size of this Box,
meaning that sufficiently reliable $T_{\rm eff}$ values must be used in order to know if
a star belongs to the Box or not. 
In this analysis we   only consider stars with a mean uncertainty  less or equal of $\pm$ 0.1 for the A(Li) and
150 K for the effective temperatures. For this, we  calculated the average between the maximum and minimum values of temperature, this is (B\_$T_{\rm eff}$ $-$ b\_$T_{\rm eff}$)/2 using the uncertainties given by  {\em Gaia} as described in  Section~\ref{Sect:DataCol}.
Let us remark that the value of an error of  150~K is near the size in $T_{\rm eff}$ of
the Box. Diving into the {\em Gaia} DR2 catalogue, which has a large
variety of $T_{\rm eff}$ errors for individual stars, we found several candidates
that are considered as the basis of our approach.

We  detect 13 stars in the Box with a mean uncertainty in $T_{\rm eff}$  of less than 150~K. Due to these uncertainties,
several of them can stay in or out of the Box.
In fact we detected 17 stars outside the Box, but due to their individual $T_{\rm eff}$ uncertainties (150~K), could get into the Box (see Table~\ref{t:box}).
We refer these stars as ``near-Box'' in the last column of Table ~\ref{t:box}.
From the total 30 stars that can participate in this enter-exit balance, there are 24 stars without detected
planets and six  planetary host stars. We note that these detected mentioned 17 stars out of the Box are all stars with $T_{\rm eff}$
values smaller than the cool limit $T_{\rm eff}$ of the Box, i.e.,   $T_{\rm eff}=$5950~K.

We also explore for eventual candidates that could enter into the Box with values larger than the hot limit of $T_{\rm eff} = $6100 K. We found seven
candidates up to maximum values of $T_{\rm eff}$ near 6300 K. Nevertheless, all of them have
mean uncertainties in $T_{\rm eff}$ larger than 150~K, then do not fulfill the conditions to
enter into the Box in this analysis.

\begin{figure}
\centering
\includegraphics[width=8.0cm]{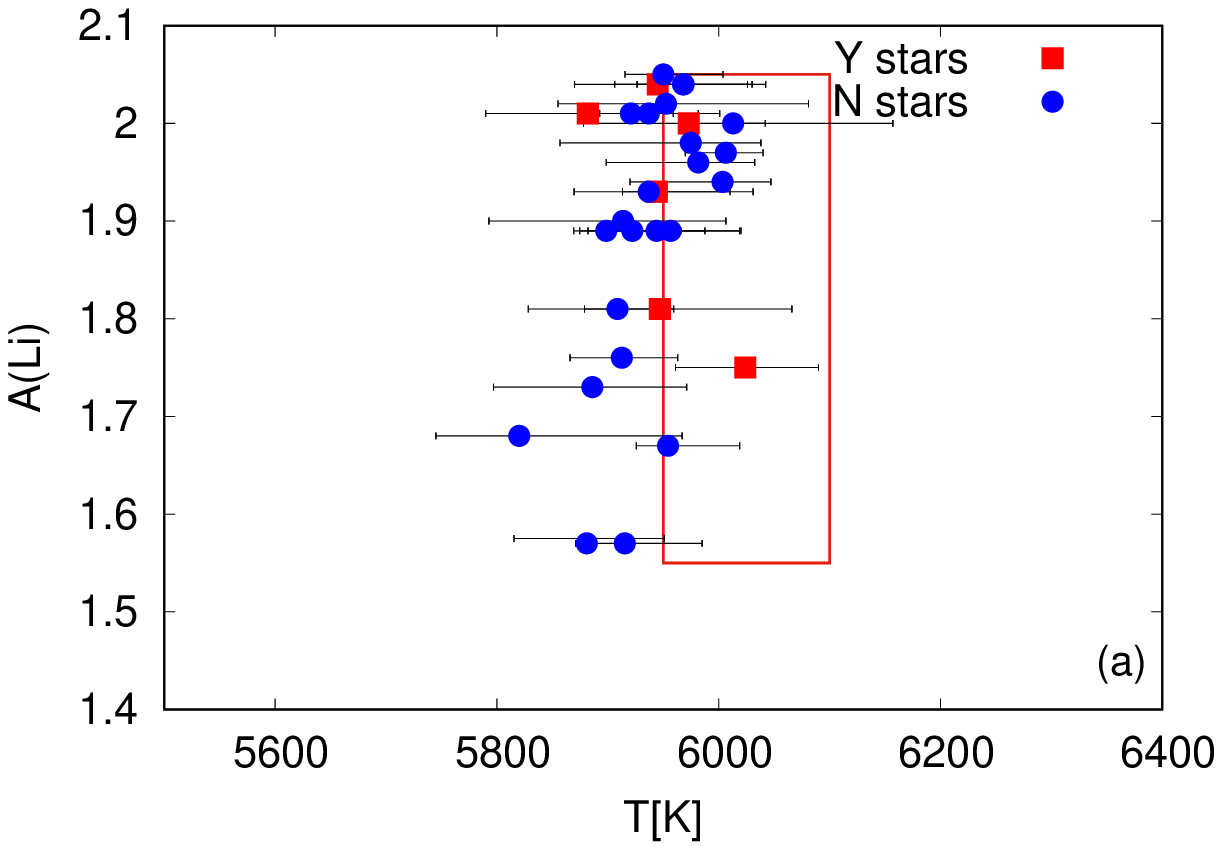}\par 
\includegraphics[width=8.5cm]{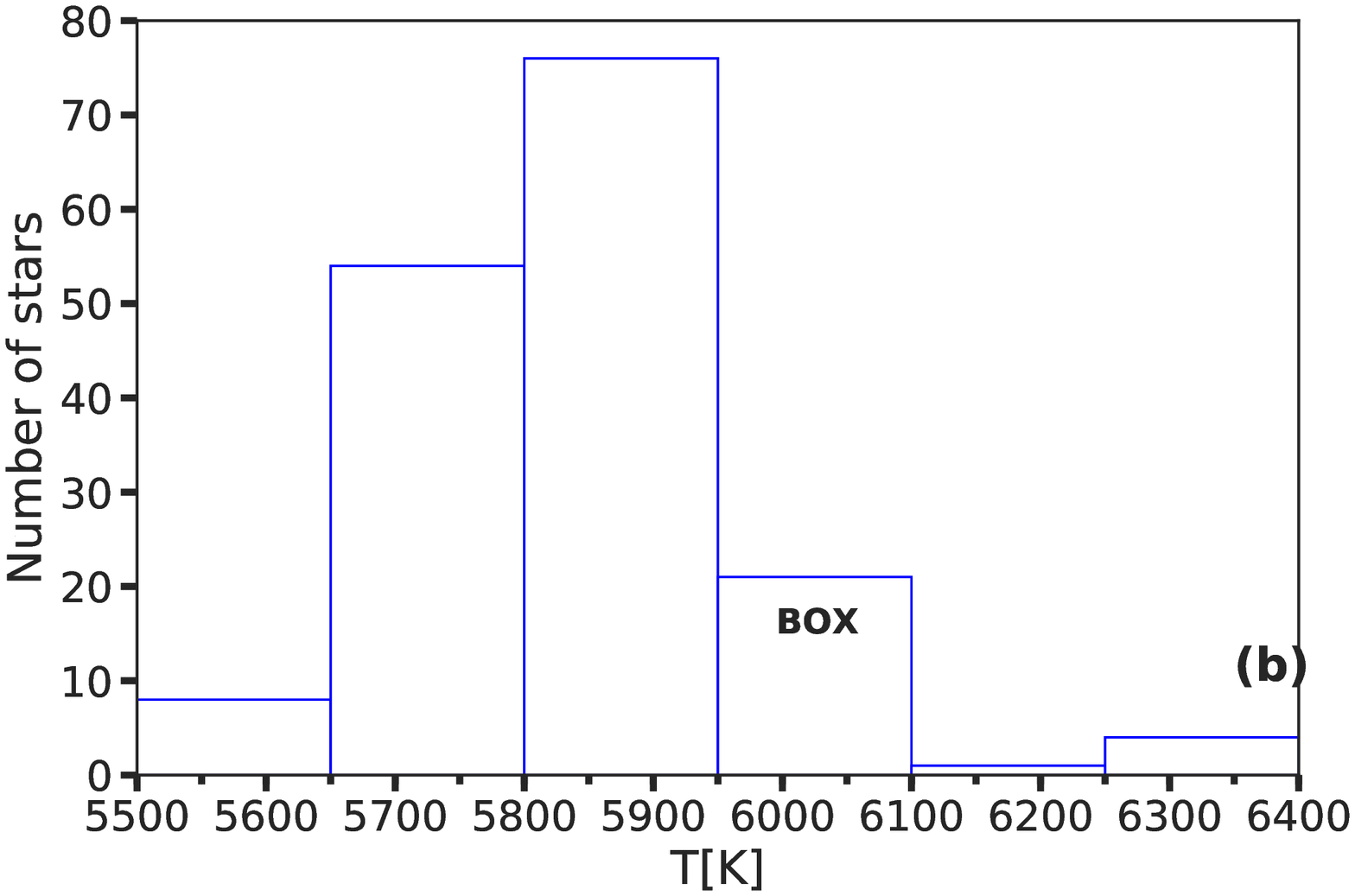}\par
\caption{ (a)A(Li) distribution  versus $T_{\rm eff}$ for the 30 stars in a near region of  the ``lithium
desert'' as defined by RA12, and represented by the green square called  Box. Stars without
detected planets (N) and stars host of planets (Y) are represented by
blue filled circles and red filled squares.
(b) Distribution in $T_{\rm eff}$  of the number of
stars in six comparative boxes around the Box for the limits 1.55 $<$ A(Li) $<$ 2.05. The
bar with the label ``Box'' represents the green square of Figure (a) corresponding to
 the region of the called lithium desert. }

\label{fig6BOXNEW}
\end{figure}

\begin{figure}
\centering

\includegraphics[width=8.5cm]{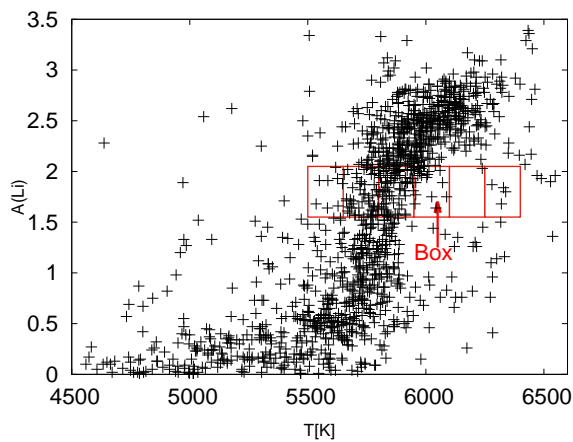}\par
\caption{A(Li) distribution  versus $T_{\rm eff}$ for all the sample of Table~\ref{t:parameters}. Six
comparative boxes of  Figure~\ref{fig6BOXNEW}b indicating the variations of the number of stars are superposed to
the Figure.  An arrow is drawn to show the location of the box corresponding
with the lithium desert. }
\label{fig7BOXNEW}
\end{figure}

When we analyse the two   stars lying in the Box detected by AG18 by means of {\em Gaia} DR2  $T_{\rm eff}$   we obtain the next: the planet host star HD 31253 () remains in all cases inside the Box, whereas the non-planetary host star HD 90422 could go out from the Box taking into account the uncertainties (see Table \ref{t:parameters}) .

Figure~\ref{fig6BOXNEW}a shows the distribution of the stars in the defined Box in the plane A(Li)-$T_{\rm eff}$
within the error limit for the effective temperatures.
It is found a proportion of six host planet stars 
against  24 non-planet host, is this a typical representative value? The answer is yes, because
this rate of 0.25 (6/24) is the same ratio that we found when considering the $\sim$ 250 host planet stars in our whole catalogue divided by the approximate value of 1070 non-planet host stars. 
By considering the uncertainties in $T_{\rm eff}$,  more stars can enter in the Box than those that can leave it.

Figure~\ref{fig6BOXNEW}b presents a histogram of effective temperature of the stars of the sample (Table \ref{t:parameters}) with 1.55 $<$ A(Li) $<$ 2.05. This is the same range in lithium abundance of the Box. For reasons of homogeneity in the comparison with the other stars in the field, the called Box in the histogram of Figure~\ref{fig6BOXNEW}b, contain a number of stars corresponding to the full sample, this is regardless the errors in $T_{\rm eff}$. For this reason, in the bin corresponding to the Box appear more stars than
those in Figure~\ref{fig6BOXNEW}a. It is clear in this histogram how the density of stars varies according to the range of temperature  considered, dropping considerably in the called Box, which is indicated in the corresponding bar.
In Figure ~\ref{fig7BOXNEW} we show the
whole map of A(Li) against $T_{\rm eff}$ ({\em Gaia}) of field stars of Table \ref{t:parameters}. The six comparative boxes
of the histogram used to see the variations of the number of stars around the Box are
superimposed to the Figure. From this point of view the ``lithium desert'' appears not to
be a real problem but more, a statistical distribution fluctuation.

\begin{table}

\caption{Parameter of stars in or near the Box of the ``lithium desert'' }
\label{t:box}
\begin{center}
\scalebox{0.8}{
\begin{tabular}{lcccccc}
\hline
\hline
Star        &$T_{\rm eff}$   &  b\_$T_{\rm eff}$  &  B\_$T_{\rm eff}$      & A(Li) &  Planet   &    Located      \\
            & (K)            &  (K)               &  (K)                   &       &     &          \\
 \hline
HD208       &  6013    &  5865      &  6157     & 2.00 $\pm$ 0.10   &  N        &    Box         \\
HD7134      &  5975    &  5858      &  6038     & 1.98 $\pm$ 0.10   &  N        &    Box         \\
HD31253     &  6025    &  5961      &  6090     & 1.75 $\pm$ 0.02   &  Y        &    Box         \\
HD36108     &  5982    &  5899      &  6033     & 1.96 $\pm$ 0.02   &  N        &    Box         \\
HD38510     &  5952    &  5855      &  6081     & 2.02 $\pm$ 0.02   &  N        &    Box         \\
HD83529     &  6004    &  5920      &  6048     & 1.94 $\pm$ 0.02   &  N        &    Box         \\
HD86081     &  5973    &  5878      &  6042     & 2.00 $\pm$ 0.08   &  Y        &    Box         \\
HD90081     &  6007    &  5970      &  6040     & 1.97 $\pm$ 0.10   &  N        &    Box         \\
HD110897    &  5968    &  5927      &  6026     & 2.04 $\pm$ 0.11   &  N        &    Box         \\
HD153627    &  5954    &  5926      &  6020     & 1.68 $\pm$ 0.02   &  N        &    Box         \\
HD165499    &  5950    &  5916      &  6005     & 2.05 $\pm$ 0.10   &  N        &    Box         \\
HD193193    &  5968    &  5906      &  6042     & 2.04 $\pm$ 0.02   &  N        &    Box         \\
HD201496    &  5957    &  5875      &  6019     & 1.89 $\pm$ 0.15   &  N        &    Box         \\
HD16382     &  5937    &  5893      &  5982     & 2.01 $\pm$ 0.10   &  N        &    Near-Box    \\
HD17865     &  5886    &  5797      &  5971     & 1.73 $\pm$ 0.12   &  N        &    Near-Box    \\
HD20407     &  5909    &  5879      &  5960     & 1.81 $\pm$ 0.12   &  N        &    Near-Box    \\
HD31527     &  5922    &  5905      &  5988     & 1.89 $\pm$ 0.02   &  N        &    Near-Box    \\
HD55575     &  5913    &  5866      &  5964     & 1.76 $\pm$ 0.10   &  N        &    Near-Box    \\
HD74957     &  5899    &  5882      &  5966     & 1.89 $\pm$ 0.10   &  N        &    Near-Box    \\
HD95128     &  5947    &  5828      &  6066     & 1.81 $\pm$ 0.04   &  Y        &    Near-Box    \\
HD97037     &  5881    &  5816      &  5950     & 1.57 $\pm$ 0.02   &  N        &    Near-Box    \\
HD114762    &  5946    &  5870      &  6030     & 2.04 $\pm$ --     &  Y        &    Near-Box     \\
HD117105    &  5937    &  5870      &  6031     & 1.93 $\pm$ 0.03   &  N        &    Near-Box    \\
HD119173    &  5914    &  5793      &  6007     & 1.90 $\pm$ 0.12   &  N        &    Near-Box    \\
HD141624    &  5915    &  5871      &  5985     & 1.57 $\pm$ 0.15   &  N        &    Near-Box    \\
HD147513    &  5883    &  5790      &  5959     & 2.01 $\pm$ 0.02   &  Y        &    Near-Box    \\
HD148816    &  5944    &  5869      &  6021     & 1.89 $\pm$ 0.02   &  N        &    Near-Box    \\
HD198089    &  5820    &  5746      &  5967     & 1.68 $\pm$ 0.05   &  N        &    Near-Box    \\
HD199289    &  5921    &  5885      &  6001     & 2.01 $\pm$ 0.10   &  N        &    Near-Box    \\
HD220689    &  5944    &  5913      &  6011     & 1.93 $\pm$ 0.02   &  Y        &    Near-Box    \\     
\hline      

\end{tabular}    }                                                                 
\tablebib{

      \item Col.1: Henry-Draper catalog name; Col.2-3-4: effective temperature taken from {\em Gaia} DR2 and their lower (b\_$T_{\rm eff}$) and upper limit (B\_$T_{\rm eff}$); Col.5: Lithium abundance; Col.6: presence of planets Yes or No, and Col.7: location in or near the  Box. 
 
    }
\end{center}
\end{table}

\section{Discussion and Conclusions}
\label{Sect:CONCLUSIONS}

This paper covers two different topics. One, in Section \ref{Sect:3rotation}, is devoted to the
lithium-rotation connection (LRC) for stars with masses between 0.8 $\mathcal{M}_\odot$ and
1.4 $\mathcal{M}_\odot$ and stellar types from F5 up to K4. The most important manifestation
of the LRC being that for very low stellar rotations, stars appear to be in general Li-poor
whereas they appear Li-rich for larger rotations.
The second topic (Section \ref{Sect:desert})
refers to the existence or not, of what is known in the literature as the 
``lithium desert'' appearing in the general distribution of the lithium abundance
A(Li) of these stars in function of effective temperatures $T_{\rm eff}$ (Figure~\ref{fignewcolors}).

We summarise  what we have learned over this study of the
lithium rotation connection.
The first new result consists of an observational determination of a critical threshold of
projected rotational velocity $v \sin i$ value which  separates slow and high stellar rotators. 
 For this purpose we used in Section \ref{Sect:3rotation} a general distribution of the Li abundance values of field
stars, with and without planets, in function of the $T_{\rm eff}$  {\em Gaia} values (Figure~\ref{fignewcolors}). To this
distribution we added a third parameter; their corresponding $v \sin i$ values. However, this
Figure does not allow us  to obtain a robust indication of this threshold velocity. A more detailed approach was
then necessary to better determine this critical velocity. 
This was done by means of the distribution of Li abundance values versus the  $v \sin i$ values adding a third axis representing other stellar parameters; that is depicted in a mosaic shown in Figure~\ref{fig3}.

A more realistic critical rotation velocity was found. 
After our analysis we conclude that those stars with $v \sin i$ $<$ 5\,km\,s$^{-1}$ (slow rotators) have different properties of age and $T_{\rm eff}$ (especially in this one) than those stars with $v \sin i$ $>$ 5\,km\,s$^{-1}$ (fast rotators), either for field stars with or without planets or cluster stars.
 In short, the critical projected rotational velocity can be considered represented by the value of 5\,km\,s$^{-1}$.
 This critical rotational velocity is representative of the lithium-rotation connection, moreover  also separates other physical parameters.

This critical velocity is important for models that study the LRC as is the case of the Li
depletion hydrodynamic model of \citet{Baraffe2017}. In this model, this
critical velocity separates the effects of penetrating plumes at the lower edge of
the convection internal zone producing the Li depletion. We should stress here,
that we also explored independently, from our data described in Section \ref{Sect:DataCol}, two
relevant catalogues of A(Li).
First, the AMBRE/Li catalogue of \citet{Guiglion2016}, which was filtered from repeated values and other inconsistencies. We
finally found for this catalogue 4927 clean actual star members of which only
2300 are dwarf FGK stars. In fact, \citet{Guiglion2016} recognise a subsample
of 2310 dwarf stars. A detailed discussion of this filtering process will be given
in our next work of these series. The second catalogue was published by \citet{Bensby2018}
which contains 515 stars. We searched, for both
catalogues, the projected rotational velocities in the literature with the same tool
as described in Section \ref{Sect:DataCol}. In both samples, we found approximately the same
threshold critical velocities as mentioned above.

Regarding our study on the age dependency of the LRC properties, we  summarise the knowledge on this subject. As mentioned in the
Introduction, \citet{Bouvier2016} have detected the presence of the LRC in a
very young open cluster of 5 Myr (NGC 2264).
It is interesting to note that this
age corresponds to the PMS stage, suggesting this way, a relation of the LRC
with the very initial stellar stages of the lives of low-mass stars. In NGC 2264,
the LRC is acting on stars with masses in the interval 0.5--1.2~$\mathcal{M}_\odot$, which is
relatively similar to the interval of masses considered here (0.8--1.4~$\mathcal{M}_\odot$). For
larger ages, \citet{Messina2016} detected the action of the LRC in the Beta
Pic association (20 Myr) but only, in a clear way, for very low mass stars
between 0.3~$\mathcal{M}_\odot$ and 0.8~$\mathcal{M}_\odot$.
For stars with masses larger than 0.8~$\mathcal{M}_\odot$ the
LRC appears only to be incipient in this association.
Several doubts arose when explaining this different behaviour, but we note that both groups are in the
general poorly known spin up stage for low mass stars, that finish at $\sim$ 30-40
Myr \citep{Bouvier2014}. The next age step of the LRC studied in the literature is
that of the open cluster Pleiades (125 Myr) and on the stellar stream Pec-Eri,
with a similar age as the Pleiades. Our contribution on the age dependency
study of LRC, refers to the second and final general rotation stage, the spin
down, also beginning at $\sim$ 30--40 Myr and lasting forever \citep{Bouvier2014}.

For all the stellar associations studied here, we first detected and eliminated all intruder members. To do
this we employed a simple chemical method by which we
considered as an intruder any star member that had an
individual metallicity value quite different from the mean
group metallicity. For this purpose we determined these
new mean group metallicities for four associations (see
Section \ref{Sect:3.2}) for which only one metallicity association
value appears in the literature. Also, considering that  the associations Tucana-Horologium, Colombus and Carina  have a similar
age of 45 Myr, this signifies that they are just at the beginning of this
general rotational braking stage. In Figure~\ref{fig3Asso} we show the behaviour of A(Li)
versus $v \sin i$ for these three associations with their genuine stellar
components, where intruder members were eliminated. In this Figure, a typical
pattern appear in which Li depleted stars are present for $v \sin i$ values less
than a critical $v \sin i$ $<$ 10\,km\,s$^{-1}$ value.

For $v \sin i$ values larger than this critical velocity, a single branch of inhibited Li depletion stars appears. This branch
continues up to very large rotational values. However, a minimum Li depletion
characteristic of these young associations, appears at $v \sin i$ $\sim$ 50\,km\,s$^{-1}$. This
kind of pattern is important, because in the next studied association, 
 AB Doradus (ABDA), an extra branch of strong Li depletion appears,
beginning at $v \sin i$ $\sim$ 30\,km\,s$^{-1}$ and continues up to large rotation values of $\sim$ 70\,km\,s$^{-1}$ as it is seen
in Figure~\ref{fig4pleiades}a. This new branch of very strong Li depletion, is
formed by low-mass K-dwarfs and their rapid Li depletion acts together 
according to their minimal masses. This behaviour of K-type dwarfs is in
agreement with the distribution of their stellar radii with $v \sin i$ as shown in
Figure~\ref{fig5NEW}. The other branch composed essentially by
only three stars of FGK-types, follows the same pattern as is the case of the
stellar groups at 45 Myr, attaining very large rotations, larger than $\sim$ 130\,km\,s$^{-1}$.
But even in this case, the only very high rapid K-dwarf rotator at $v \sin i = $140\,km\,s$^{-1}$, is the only one that
depletes Li.
Let us note that this double branch depletion pattern in ABDA, it is
also found in the Pleiades (see Figure \ref{fig4pleiades}b).
This similarity agrees with the fact
that both, the ABDA and the Pleiades, have not also the same age around
120 Myr but also that both stellar groups  have a common origin \citep{Ortega2007}.

We understand that the formation of a second depletion branch in groups at
120 Myr, is a consequence of the general diminishing of stellar rotation for
that age, during the general spin down era. We remark that the birth of the first
depletion branch in ABDA at $v \sin i$ $\sim$ 30\,km\,s$^{-1}$, corresponds also to the
minimum Li depletion of the Pleiades as it is seen in Figure \ref{fig4pleiades}. This value is smaller than the
value of $v \sin i$ $\sim$ 50\,km\,s$^{-1}$ of the mentioned minimum Li depletion in the
younger associations. We note, however, that for the open cluster Alpha Per
with an age of (50-70 Myr) which is relatively intermediate between of our
considered younger associations (45 Myr) and that of ABDA (120 Myr), the
Li minimum also appears at $v \sin i$ $=$ 50\,km\,s$^{-1}$ but even more, at this velocity a
double branch is   formed similarly as is the case of ABDA   \citep[see Figure 3 in][]{Balachandran88}.
In this Figure it can be seen that the LRC pattern in
Alpha Per, at this intermediate age, shows a mixture of properties between
the associations at 45 Myr and 120 Myr mentioned before.

To explore the behaviour of the LRC for larger ages, we considered the open cluster NGC
1039 with an age of 250 Myr. Following the classical publication of this cluster
by \citet{Jones1997}, we plot the A(Li) versus $v \sin i$ values (plot not shown in this
paper) for members with more than 75\% member probability. The
corresponding minimum of Li depletion corresponds to near 17 km/s.
Separately, we confirmed this same value of $\sim$ 17\,km\,s$^{-1}$, by plotting for
members of NGC 1039 with more than 90\% member probability, by using the
A(Li) values of  \citet{SestitoRandich2005} and $v \sin i$  values compiled and described in Section \ref{Sect:DataCol}. We conclude that in a period going from the age of 45 Myr of the younger associations considered here, up to the age of
250 Myr of NGC 1039, (i.e.  a 200 Myr interval), their minimum Li depletion
points have shifted from the corresponding $v \sin i$ value of $\sim$ 50\,km\,s$^{-1}$ at 45 Myr
to somewhat less than 20\,km\,s$^{-1}$ for 250 Myr. This shift appears also to be a
direct consequence of the general spin down of the stars rotation behaviour.
Other conclusion of this work, is that K-dwarf stars are the first to deplete their
Li in these early stages in the main-sequence at large rotation velocities, due
to their important convective zones. This behaviour is also due to the general
spin down of low mass stars. We can try to quantify this behaviour by
estimating the mean $v \sin i$ values for our studied stellar moving groups with
ages at 45 Myr and 120 Myr. These approximate mean values of $v \sin i$ are
respectively equal to 32\,km\,s$^{-1}$ and 20\,km\,s$^{-1}$. Their ratio 32/20 = 1.6 can be
compared with that expected in the general theoretical braking rotational
curves, for slow and fast rotators   \citep[see Figure 7 in][]{Bouvier2014}. For
these ages, the ratios are 1.6 and $\sim$ 2.0 respectively, which is in agreement
with the observed values of our chosen associations.

Section~\ref{Sect:desert}  was devoted entirely   to the study of what was considered in the
literature as the  ``lithium desert''.
In a first sight, this region shows an anomalous absence of stars which also
presented an unsolved problem, regarding as which is the physical
mechanism that provokes such a radical Li depletion. In reality, by means of
new $T_{\rm eff}$ values obtained from {\em Gaia} DR2, we found that the ``lithium desert''
appears to be a false problem as we detected in this work 30 stars that are in
the box defined above or very near to it. Due to the  uncertainties of the
{\em Gaia}  $T_{\rm eff}$ values, from these 30 objects, 13 stars in the box can remain or
leave the box and nearby 17 stars can enter into the box. In any case, the box   representing the ``lithium desert'' probable contain stars.
A population
test was made  to see if a possible inference of the actual number of stars of
the Box is reasonable. All this in comparison with the whole nearby stars population
in that A(Li) interval. We found that approximately 15 stars can occupy the region of
what was called the ``lithium desert''. This shows that the considered past absence of stars
was more likely a result of an statistical fluctuation distribution effect.
\section*{Acknowledgements}

The authors thank the anonymous referee for the constructive comments and helpful insights. 
This work has made use of {\tt SVODiscTool}, developed by the Spanish Virtual Observatory (Centro de Astrobiolog\'{\i}a (CSIC-INTA), Unidad de Excelencia Mar\'{\i}a de Maeztu), a project supported by the Spanish State Research Agency (AEI) through grants AyA2017-84089 and MDM-2017-0737. C.CH acknowledges support from SECYT/UNC
and CONICET. FLA and R de la R. acknowledge support from the Faculty of the European Space Astronomy Centre (ESAC) - Funding references 569 and 570,  respectively. FLA would like to thank the technical support provided by A. Parras (CAB), Dr. J. A. Prieto (UCLM) and MSc J. G{\'o}mez Malag{\'o}n. Authors also thank to Leo Girardi for providing the {\tt PARAM} code. SFR acknowledges support from a Spanish postdoctoral fellowship `Ayudas para la atracci{\'o}n del talento investigador. Modalidad 2: j{\'o}venes investigadores, financiadas por la Comunidad de Madrid' under grant number 2017- T2/TIC-5592. SRF acknowledges financial support from the Spanish Ministry of Economy and Competitiveness (MINECO) under grant number AYA2016-75808-R, AYA2017-90589-REDT and S2018/NMT-429, and from the CAM-UCM under grant number PR65/19-22462 and from the CAM-UCM under grant number PR65/19-22462.
CC acknowledges financial support from the Spanish Ministry of Science and Innovation through grants AYA2016-79425- C3-1/2/3-P and BES-2017-080769

\end{document}